\documentclass{aa}
\usepackage[colorlinks, citecolor=blue]{hyperref}
\usepackage{ctable}
\usepackage{graphicx,epsfig,fancyhdr,psfig,rotating,amsmath,epsf,txfonts,natbib,epstopdf,multirow}

\def \kms {{ \rm km\;s$^{-1}$}}

\begin{document}

\title{Coronal hole boundaries at small scales: III.  EIS and SUMER views}

\author{M.S. Madjarska\inst{1,2}, Z. Huang\inst{1}, J.G. Doyle\inst{1} \and S. Subramanian\inst{1}}

\offprints{madj@arm.ac.uk}
\institute{Armagh Observatory, College Hill, Armagh BT61 9DG, N. Ireland
\and UCL-Mullard Space Science Laboratory, Holmbury St Mary, Dorking, Surrey, RH5 6NT, UK}
 \date{Received date, accepted date}

\abstract
{We report on the plasma properties of  small-scale transient  events identified in the quiet Sun, coronal holes and their boundaries.}
{We aim at deriving the physical characteristics of events which were identified as small-scale transient  brightenings 
in XRT images.}
{We use  spectroscopic co-observations  from SUMER/SoHO and EIS/Hinode combined with high cadence imaging data from XRT/Hinode. We 
measure Doppler shifts using single and multiple Gauss fits of transition region and coronal lines as well as electron densities and 
temperatures. We combine co-temporal imaging and spectroscopy to separate brightening expansions from plasma flows.}
{The transient brightening events  in coronal holes and their boundaries were found to be very dynamical 
producing high density outflows at large speeds. Most of these events represent X-ray jets from pre-existing or newly emerging  
coronal bright points at X-ray temperatures. The average electron density of the jets is log$_{10}{N_e}$ $\approx$8.76~cm$^{-3}$ 
while in  the flaring site it is  log$_{10}{N_e}$ $\approx$9.51~cm$^{-3}$. The jet temperatures reach a maximum of 2.5~MK but in the majority of the cases the 
temperatures do not exceed 1.6~MK. The footpoints of jets have temperatures of a maximum of 2.5~MK though in a single event scanned a minute after the flaring the measured temperature  was 12~MK. The jets are produced by multiple microflaring in the transition region and corona. Chromospheric emission was only detected in their footpoints and was only associated with downflows. The Doppler shift measurements in the quiet Sun transient brightenings confirmed that these events do not produce jet-like phenomena. The plasma flows in these phenomena remain trapped in closed loops. }
{We can conclude that the dynamic day-by-day and even
hour-by-hour small-scale evolution of coronal hole boundaries
reported in paper I is indeed related to coronal bright points.
The XRT observations reported in paper II  revealed that these changes are associated with the dynamic evolution of coronal
bright points producing multiple jets during their lifetime
until their full disappearance. We 
demonstrated here through spectroscopic EIS and SUMER co-observations combined with high-cadence imaging information 
that the co-existence of open and closed magnetic fields results  in multiple energy depositions which propel 
high density plasma along open magnetic field lines. We conclude from the physical characteristics  obtained in this study that X-ray jets are
 an important  candidate for the source of the slow solar wind. This, however, does not exclude the possibility that these jets are also 
 the microstreams observed in the fast solar wind as recently suggested.}

\authorrunning{Madjarska, M. S. et al.}
\titlerunning{Coronal hole boundaries at small scales}

\maketitle

\section{Introduction}
\label{intro}
Coronal Holes (CHs) are regions on the Sun dominated by open magnetic fields. They are seen with 
reduced emission in spectral lines formed at coronal temperatures and are identified 
as the source regions of the fast solar wind with velocities of $\sim$800~\kms\ \citep{1973SoPh...29..505K}. 
 CHs form in both polar and equatorial regions. The latter are often connected with the polar CHs with a channel of open magnetic field
 and are called equatorial extensions of polar coronal holes (EECHs). EECHs 
  exhibit a more rigid rotation \citep{1975SoPh...42..135T} 
with respect to the typical differential rotation at photospheric levels. Thus, it is believed that interchange 
magnetic reconnection happens excessively  between the open (coronal hole's) and closed (quiet Sun) magnetic field lines.  
This was suggested to play an  important role in the generation of the slow solar wind flow 
\citep{1998ApJ...498L.165W, 2004ApJ...612.1171W}. Evidence of dynamic processes taking place at coronal hole boundary (CHB) 
regions were first provided from spectroscopic observations by  \citet{2004ApJ...603L..57M}.
It has been found that events described by non-Gaussian profiles of transition region spectral 
lines, e.g. in N~{\sc iv}~765.15~\AA\ and Ne~{\sc viii}~770.43~\AA, were abundant along the boundary of  
an EECH. Similar results were found at a polar coronal hole 
boundary by \citet{2006A&A...446..327D}  in O~{\sc vi}~1031.93~\AA. More details on the background of 
CHs and coronal hole boundaries can be found in \citet[][hereafter paper I]{2009A&A...503..991M} and 
\citet[][hereafter paper II]{2010A&A...516A..50S}.
 
\citet{2009A&A...503..991M} showed that, although isolated equatorial CHs and EECHs maintain their general shape during several solar rotations, a closer look 
at their day-by-day and hour-by-hour evolution demonstrates significant dynamics. Using Extreme-ultraviolet Imaging Telescope (EIT)/SoHO 195~\AA\ 
and TRACE 171~\AA\ observations, they found that evolution of small-scale loops, i.e. 
coronal bright points (BPs), led to small-scale changes of the CH boundaries. The disappearance of BPs was associated with an 
expansion of the CHs while the appearance of BPs was followed by contraction of the CHs.

Since the launch of the Hinode satellite, coronal holes are seen in observations from the X-ray Telescope \citep[XRT,][]{2007SoPh..243...63G} as 
highly structured and dynamically evolving at small scales as never before. We analysed XRT images taken with
the Al$\_$poly filter (paper II) studying  transient brightenings in CH, CHB and quiet Sun regions. We 
found that 70\% of the brightening events in CHs and at CH boundaries appear as loop expulsions  and/or 
collimated outflows while only 30\% of the events in the quiet Sun showed 
possible flows/outflows. We also found that the ejected plasma during outflow events always originated 
from pre-existing or newly emerging coronal bright points at X-ray temperatures. \citet{madj2010} studied in great detail  an  X-ray jet  
based on  unique imager and spectrometer co-observations. The author found that the outflow is composed both of BP loop 
expulsion and collimated outflow with several energy depositions during the  lifetime of the event. Hot plasma was seen to rise earlier 
than the cooler plasma. The spectroscopic study revealed two main flows for the jet, a fast outflow with Doppler velocities 
around 300~\kms\ and a downflow guided by closed 
magnetic field lines with Doppler velocities of 25 to 50~\kms\ at transition region temperatures and 
150~\kms\ at coronal temperatures. They also established that multiple energy depositions happened 
during the jet. 

X-ray jets were first discovered in observations by the Soft X-ray 
Telescope (SXT) onboard \textit{Yohkoh}. They 
are  associated with BPs \citep{1992PASJ...44L.173S,1994ApJ...431L..51S, 2010ApJ...710.1806D, 1998ApJ...508..899W,2010A&A...516A..50S}. In 
a statistical study, \citet{1996PASJ...48..123S} concluded that they have an average length of a few 
$10^4$ to $4\times10^5$~km, widths of $5\times10^3$ to $10^5$~km, velocities of 10 to 1000~\kms\ with 
an average velocity of 200~\kms\ and lifetime of up to 10 hours. Although 68\% of X-ray jets were 
discovered in or near active regions, they can also be seen in the quiet Sun and CHs. Furthermore, 
\citet{2000ApJ...542.1100S} analysed the  physical parameters of 16 X-ray jets using imaging data 
from the soft X-ray telescope onboard the Yohkoh satellite. They derived the temperature of the jets and  their flaring footpoint as 3--8~MK. The electron density of the jets
was found to be  0.7--$4\times10^9$~cm$^{-3}$ while for the flaring site  
2.4--$10.0\times10^9$~cm$^{-3}$. These events are also observed in 
the ultraviolet \citep{1999SoPh..190..167A, 2005ApJ...623..519K, madj2010} and white-light 
\citep{2010SoPh..264..365P}.  
\citet{2010SoPh..264..365P} analysed over 10\,000 events observed in white-light by STEREO/SECCHI--COR1, 
and found that little more than 70\% of the white-light jets were associated with EUV jets.  Usually, X-ray jets are  associated with small-scale flares in their footpoints 
\citep{1992PASJ...44L.173S}. Hence, magnetic reconnection is suggested as their formation mechanism 
\citep{1992PASJ...44L.173S, 1995Natur.375...42Y, 2000ApJ...542.1100S, 2008ApJ...673L.211M,2010ApJ...714.1762P}.

The occurrence of non-Gaussian profiles in spectral lines from ions formed at transition region temperatures was 
named as an `explosive event' (EEs)  with no imaging information of what  phenomenon  actually produces these line profiles.
These events are studied for the past three decades using observations obtained first by the
Naval Research Laboratory High Resolution Telescope
and Spectrometer  (HRTS) and, later, the Solar Ultraviolet Measurements
of Emitted Radiation (SUMER) spectrograph onboard the Solar and Heliospheric Observatory (SoHO). 
Explosive events were found to have an average lifetime of 200~s and a size of 4\arcsec\ -- 5\arcsec\  
determined by the width of the blue- and red-shifted emission along a spectrometer slit. EEs show 
Doppler velocities of up to 200~\kms, predominantly blue-shifted, and  were found at the
boundaries of the super-granulation cells. The main characteristics of EES  were studied 
by \citet[][and the references therein]{1983ApJ...272..329B, 1989SoPh..123...41D, 1991ApJ...370..775P,  2002A&A...382..319M, 2003A&A...403..731M}. 
It has been suggested that EEs are the spectroscopic signature of magnetic reconnection happening 
in the transition region \citep{1994AdSpR..14...13D,1997Natur.386..811I}. In recent years, simultaneous 
imaging and spectroscopic observations helped to shed more light on the nature of this phenomenon.  
EEs were found to be produced by a siphon flow in a small-scale loop \citep{2004A&A...427.1065T}. They also 
resulted from the up- and down-flows in a surge  \citep{2009ApJ...701..253M}.

\begin{table}[ht!]
\centering
\caption{The  spectral lines registered with EIS. The comment `b' means that the spectral line is 
blended. The spectral lines with an asterisk were taken during the January 2009 observations.}
\begin{tabular}{l l c}
\hline\hline
Ion & Wavelength (\AA) & logT$_{max}$ (K) \\
\hline
He~{\sc ii}$^*$ & 256.32b & 4.7 \\
O~{\sc v}$^*$ & 192.90 & 5.4\\
O~{\sc vi}$^*$ & 184.12 & 5.5\\
Mg~{\sc vi}& 270.39b & 5.7\\ 
Mg~{\sc vii} & 278.39 & 5.8\\
& 280.75 & 5.8 \\     
Si~{\sc vii}$^*$ & 275.35 & 5.8 \\
Fe~{\sc viii}$^*$ &185.21&5.8\\
Fe~{\sc x}$^*$ & 184.54&6.0  \\
Fe~{\sc xi} & 188.23 &6.1 \\
Si~{\sc x} & 258.37 &6.1 \\
Si~{\sc x}$^*$  & 261.04 &6.1  \\
Fe ~{\sc xii}$^*$ & 195.12 &6.1\\
Fe~{\sc xiii}$^*$&202.04&6.2\\
&203.82&6.2\\
Fe~{\sc xiv}$^*$ &274.20b &6.3 \\
Fe~{\sc xv}$^*$ & 284.16 &6.3 \\
Fe~{\sc xvi}&262.98&6.4\\
Ar~{\sc xv}&293.69&6.6\\
Fe~{\sc xxiii}&263.76&7.1\\
\hline
\end{tabular}
\label{tb1}
\end{table}

\begin{table}[ht!]
\centering
\caption{The SUMER spectral lines. The expression `/2' means that the spectral
 line was observed in second order. The comment `b' means that the spectral line is 
 blended with a  close-by line. }
\begin{tabular}{l l c}
\hline\hline
Ion & Wavelength (\AA) & logT$_{max}$ (K)\\
\hline
N~{\sc v} & 1238.82 & 5.3 \\
C~{\sc i} & 1248.00 & 4.0 \\
& 1248.88 & \\
C~{\sc i} & 1249.00 & 4.0 \\
O~{\sc iv}/2 & 1249.24b & 5.2 \\ 
Si~{\sc x}/2 & 1249.30b & 6.1  \\
C~{\sc i} & 1249.41 & 4.0 \\     
Mg~{\sc x}/2 & 1249.90 & 6.1 \\
O~{\sc iv}/2 &1250.25b&5.2\\
Si~{\sc ii} & 1250.09&4.1   \\
Si~{\sc ii} & 1250.41 &4.1  \\
C~{\sc i} & 1250.42b & 4.0 \\
S~{\sc ii} & 1250.58 &4.2 \\
Si~{\sc ii} & 1251.16 &4.1  \\
C~{\sc i} & 1251.17b &4.0\\
O~{\sc iv}/2&1251.70&5.2\\
& 1251.78 &\\
Si~{\sc i} & 1256.49&4.0\\
Si~{\sc i} & 1258.78 &4.1 \\
S~{\sc ii} & 1259.53b &4.2 \\
O~{\sc v}/2&1259.54&5.4  \\
Si~{\sc ii}&1260.44&4.1\\
Si~{\sc iv}&1393.78 &4.9\\
\hline
\end{tabular}
\label{tb2}
\end{table}

\begin{table*}
\centering
\caption{Description of  the EIS observations used in this study.}
\begin{tabular}{c c c c c c}
\hline\hline
Date &Observing period& \multicolumn{3}{c}{Heliospheric coordinates (min -- max)}\\
(DD/MM/YY)&(UT)&\multicolumn{2}{c}{Solar X (arcsec)}&Solar Y (arcsec)& Number of rasters\\
&&Big Raster&Small Raster&\\
\hline
09/11/07&06:35 $\rightarrow$ 14:50&$-$545.2 $\rightarrow$ $-$425.2&$-$499.3 $\rightarrow$ $-$475.3&$-$421.1 $\rightarrow$ $-$24.1& 1 big, 29 small\\
12/11/07&01:17 $\rightarrow$ 10:43&$-$40.8 $\rightarrow$ 79.2   &8.2 $\rightarrow$ 32.2&$-$432.0  $\rightarrow$ $-$35.0& 1 big, 35 small\\
14/11/07&00:20 $\rightarrow$ 10:48&208.9 $\rightarrow$ 328.9  &261.0 $\rightarrow$ 285.0&$-$382.6 $\rightarrow$ 14.4& 1 big, 40 small\\
16/11/07&18:07 $\rightarrow$ 23:47&729.1 $\rightarrow$ 849.1  &784.3 $\rightarrow$ 808.3&$-$341.3 $\rightarrow$ 55.7& 1 big, 22 small\\
10/01/09&11:30 $\rightarrow$ 17:27&\multicolumn{2}{c}{$-$100.0 $\rightarrow$ $-$30.0}&$-$146.7 $\rightarrow$ 101.3& 14\\
13/01/09&11:22 $\rightarrow$ 17:41&\multicolumn{2}{c}{$-$101.2 $\rightarrow$ $-$31.2}&$-$147.0 $\rightarrow$ 101.0 & 15\\
\hline
\end{tabular}
\label{tb3}
\end{table*}

\begin{table*}
\centering
\caption{Description of the SUMER observations used in this study.}
\begin{tabular}{c c c c c}
\hline\hline
Date &\multicolumn{2}{c}{Observing period (UT)}& \multicolumn{2}{c}{Heliospheric coordinates}\\
(DD/MM/YY)&Sit-and-stare&Raster&Solar X (arcsec)&Solar Y (arcsec)\\
\hline
\multirow{2}{*}{09/11/07}&07:01 $\rightarrow$ 08:41&08:43 $\rightarrow$ 08:58&\multirow{2}{*}{$-$501.6}&\multirow{2}{*}{$-$339.1 $\rightarrow$ $-$39.1}\\
 &11:01 $\rightarrow$ 12:42&12:43 $\rightarrow$ 12:58&&\\
 \hline
\multirow{2}{*}{12/11/07}&01:35 $\rightarrow$ 03:16&03:17 $\rightarrow$ 03:31&\multirow{2}{*}{$-$5.6}&\multirow{2}{*}{$-$345.0 $\rightarrow$ $-$45.0}\\
&05:34 $\rightarrow$ 07:14&07:16 $\rightarrow$ 07:31&&\\
\hline
\multirow{3}{*}{14/11/07}&01:01 $\rightarrow$ 02:41&02:43 $\rightarrow$ 02:56&\multirow{3}{*}{245.2}&\multirow{3}{*}{$-$300.6 $\rightarrow$ $-$0.6}\\
&02:58 $\rightarrow$  04:39&04:40 $\rightarrow$ 04:55&&\\
&07:01 $\rightarrow$ 08:42&08:43 $\rightarrow$ 11:02&&\\
\hline
\multirow{2}{*}{16/11/07}&18:01 $\rightarrow$ 19:41&19:42 $\rightarrow$ 19:57&\multirow{2}{*}{750.5}&\multirow{2}{*}{$-$259.3 $\rightarrow$ 40.7}\\
&22:16 $\rightarrow$ 23:41&23:43 $\rightarrow$ 23:58&&\\
\hline
\end{tabular}
\label{tb4}
\end{table*}

Here, we present a follow-up study on coronal hole boundary evolution, namely a spectroscopic 
analysis of some of the brightening  events identified in the XRT images from November 2007 and January 2009 studied in paper~II.  In Sect.~2 we describe the observations. 
The data analysis and results are given in Sect.~3. 
The discussion and conclusions  are outlined in Sect.~4.

\section{Observations}
\label{sect2}

\begin{figure*}[!ht]
  \centering
  \includegraphics[width=19cm]{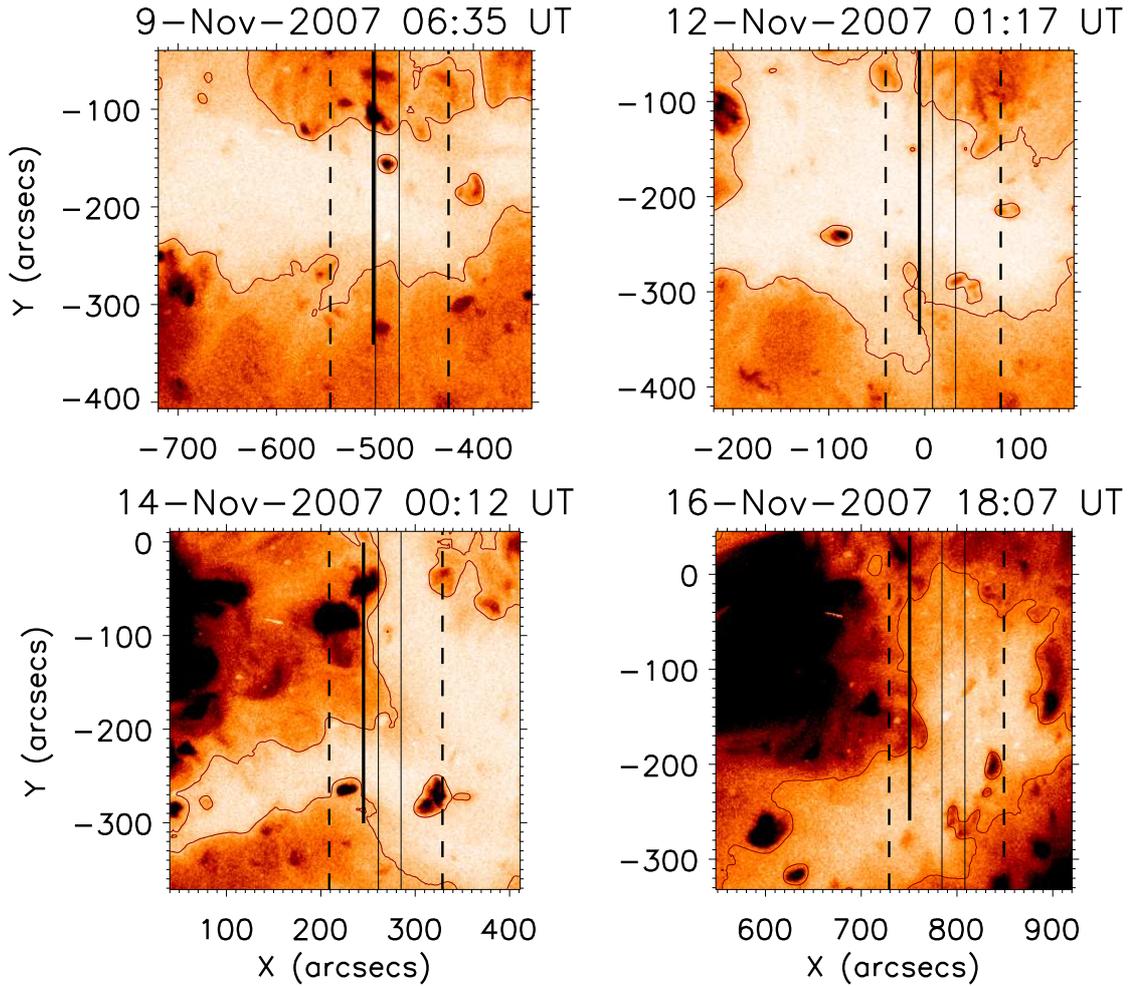}
   \caption{The XRT images of the four days of coronal hole observations. The dashed lines show the 
   large EIS raster field-of-view while the thin solid lines denote the small raster FOV. The SUMER 
   slit-and-stare position is shown with a thick solid line. Contours of XRT images outline the coronal 
   holes as defined in paper~II.}
   \label{fig1}
\end{figure*}

\begin{figure*}[!ht]
\vspace{7cm}
  \centering
    \includegraphics{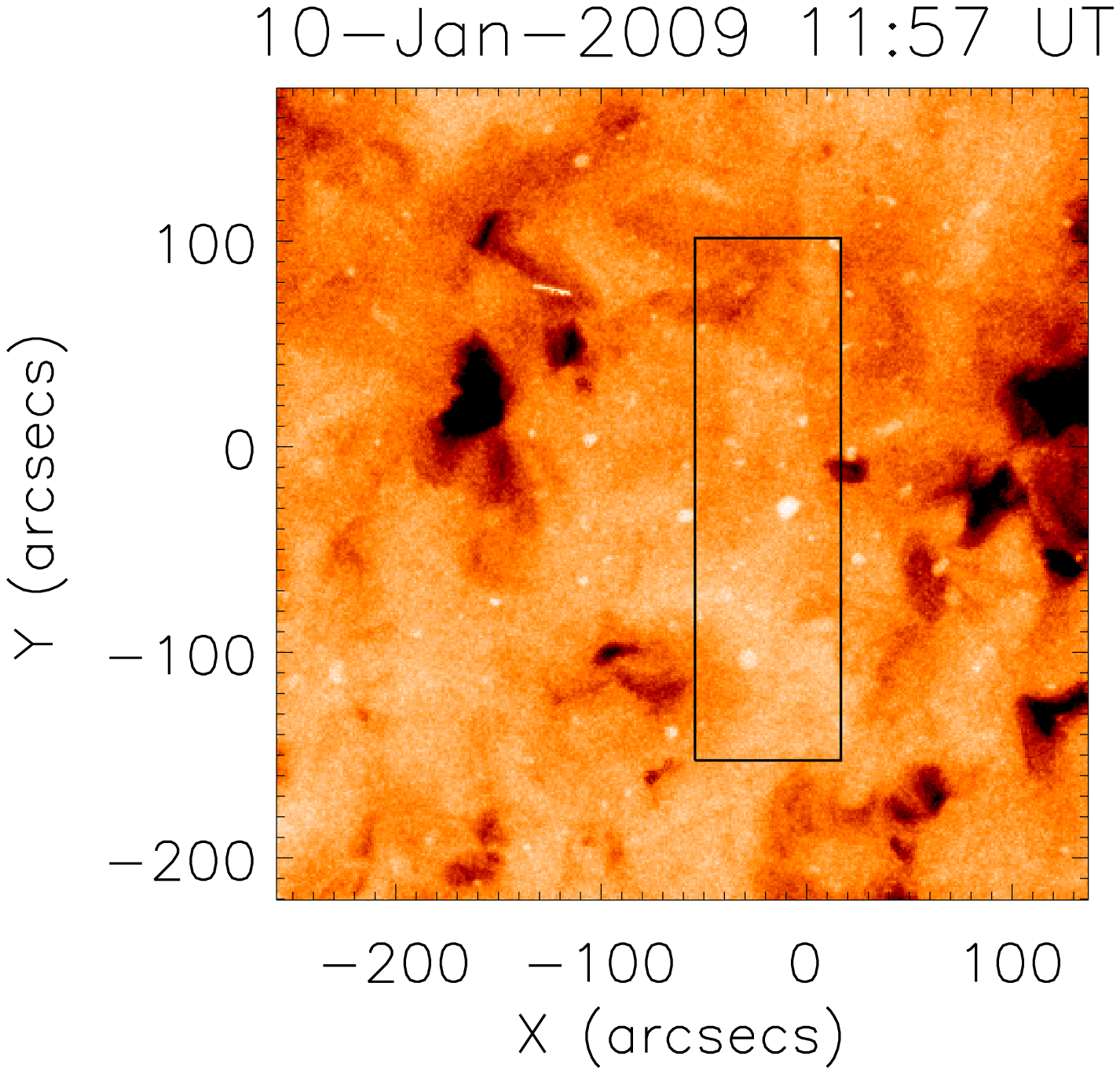}
\includegraphics{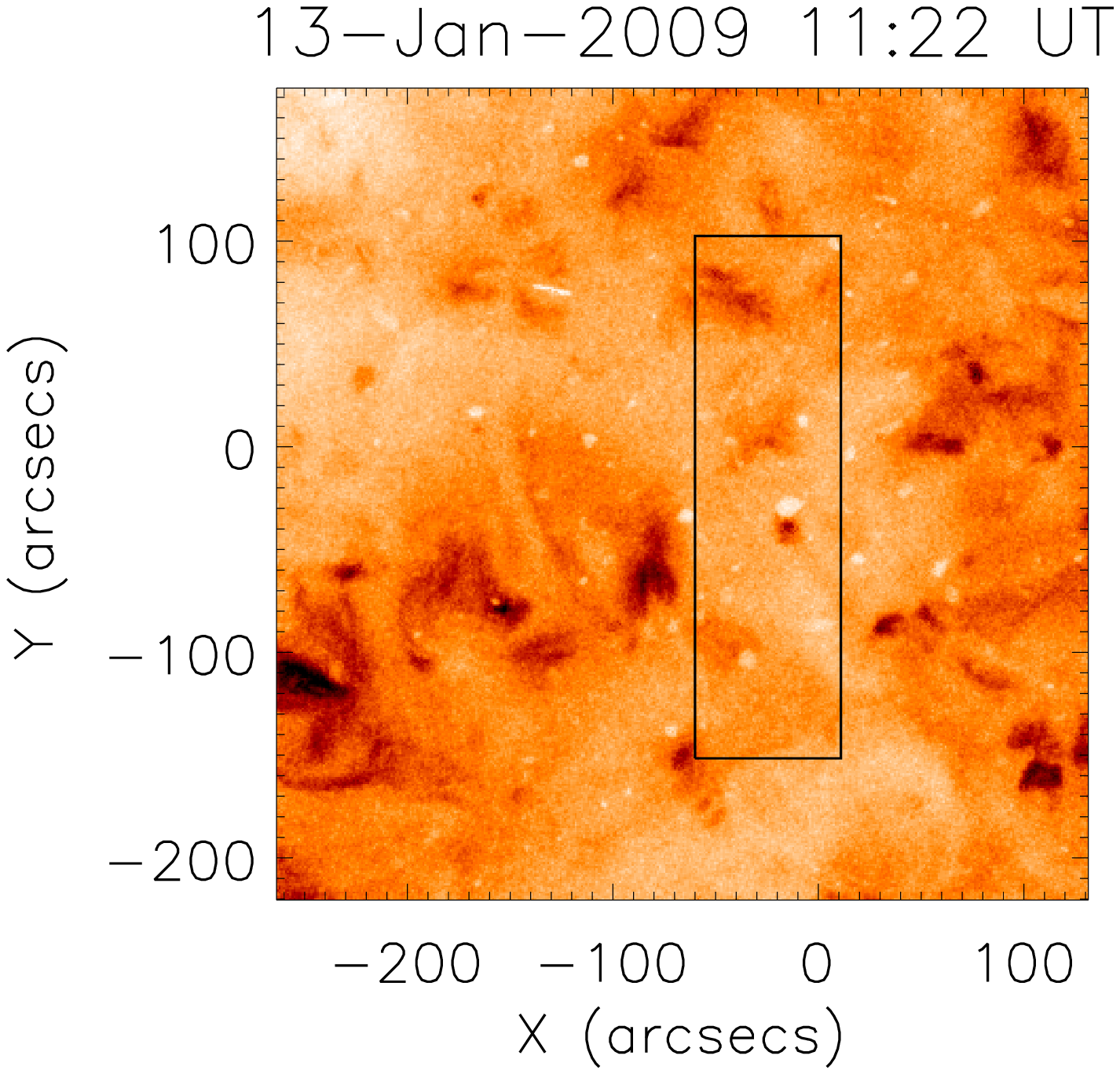}
   \caption{The XRT images taken on 2009 January 10 (left) and 13 (right) in the quiet Sun with  
    the field-of-view of the EIS rasters overplotted.   \label{fig2}}
\end{figure*}

The observations used for the present study were taken on November 2007 during  a dedicated 
multi-instrument Hinode and SoHO campaign with the SUMER/SoHO and Extreme-ultraviolet Imaging Telescope (EIS)/Hinode.  
 An equatorial coronal hole was tracked  from the East 
to the West limb during four days. Details on the registered EIS and SUMER spectral 
lines can be found in Table~\ref{tb1} and Table~\ref{tb2}, respectively. Details on the 
observing times and pointing of the SUMER and EIS instruments
are given in Table~\ref{tb3} and Table~\ref{tb4}. We also analysed data taken in the  quiet Sun 
in January 2009 (all the information on these data is also included in Tables~1--4). 
The  field-of-views (FOVs)  of XRT, EIS and  SUMER are shown in Figs.~\ref{fig1} and \ref{fig2}. The SUMER 
and EIS data reduction and alignment were done as described in \citet{madj2010}. 

%%%%%%%%%%%%%%% Fig 1 %%%%%%%%%%%%%%%%%%%

%%%%%%%%%%%%%%% Fig 2 %%%%%%%%%%%%%%%%%%%

\subsection{EIS}

The EIS \citep{2007SoPh..243...19C} study was specially designed to provide the best possible 
coverage of spectral lines with formation temperatures  (logT$_{max}$) from 4.7~K to 7.1~K. First, a raster 
of 120\arcsec $\times$ 512\arcsec\ was obtained, followed by small rasters of 
24\arcsec~$\times$~512\arcsec.  The 2\arcsec\ slit with an exposure time of 60~{\rm s} was used. During the 
January 2009 run the rasters had a size of 70\arcsec~$\times$~248\arcsec\ and were taken with a 40~{\rm s}  
exposure time. The Doppler shift maps were derived from a single Gauss fit with the reference line obtained 
as an average over the entire raster after all instrumental effects were removed. The Doppler velocities 
measured here should be considered as indicative only of the range of the Doppler shift rather than absolute 
values as EIS does not have an absolute calibration although for some type of observations Doppler measurements close 
to their absolute values can be obtained \citep{2012ApJ...744...14Y}.  

\subsection{SUMER}

The SUMER \citep{1995SoPh..162..189W,
1997SoPh..170..105L} observations were made with a slit size of 
1\arcsec$\times$300\arcsec. The detector B was exposed for 60~{\rm s} while the spectrometer was 
observing in a sit-and-stare mode. The time sequence was followed by a raster with a
size of 60\arcsec $\times$ 300\arcsec\  made only in the O~{\sc v}~629.77 \AA\ and  
Si~{\sc i}~1256.49~\AA\ lines and a 30~{\rm s} exposure time. The observations were compensated for the solar rotation. Five 
spectral  windows were transferred to the ground  each with a size of 50 spectral pixels $\times$ 300 spatial 
pixels. During the November 9 observations the detector read-outs were shifted by around 20 spectral pixels. As a 
result the O~{\sc v} and N~{\sc v} lines have part of their blue wing cut which limited the line blueshift 
measurements to only 92~\kms.  From all lines only  O~{\sc v}~629.77~\AA\  (in second order of reflection) was taken on the bare 
part of the detector. 

%%%%%%%%%%%%%%% Fig 3 %%%%%%%%%%%%%%%%%%%

 \begin{figure*}[!ht]
  \centering
  \vspace{11cm}
  \includegraphics{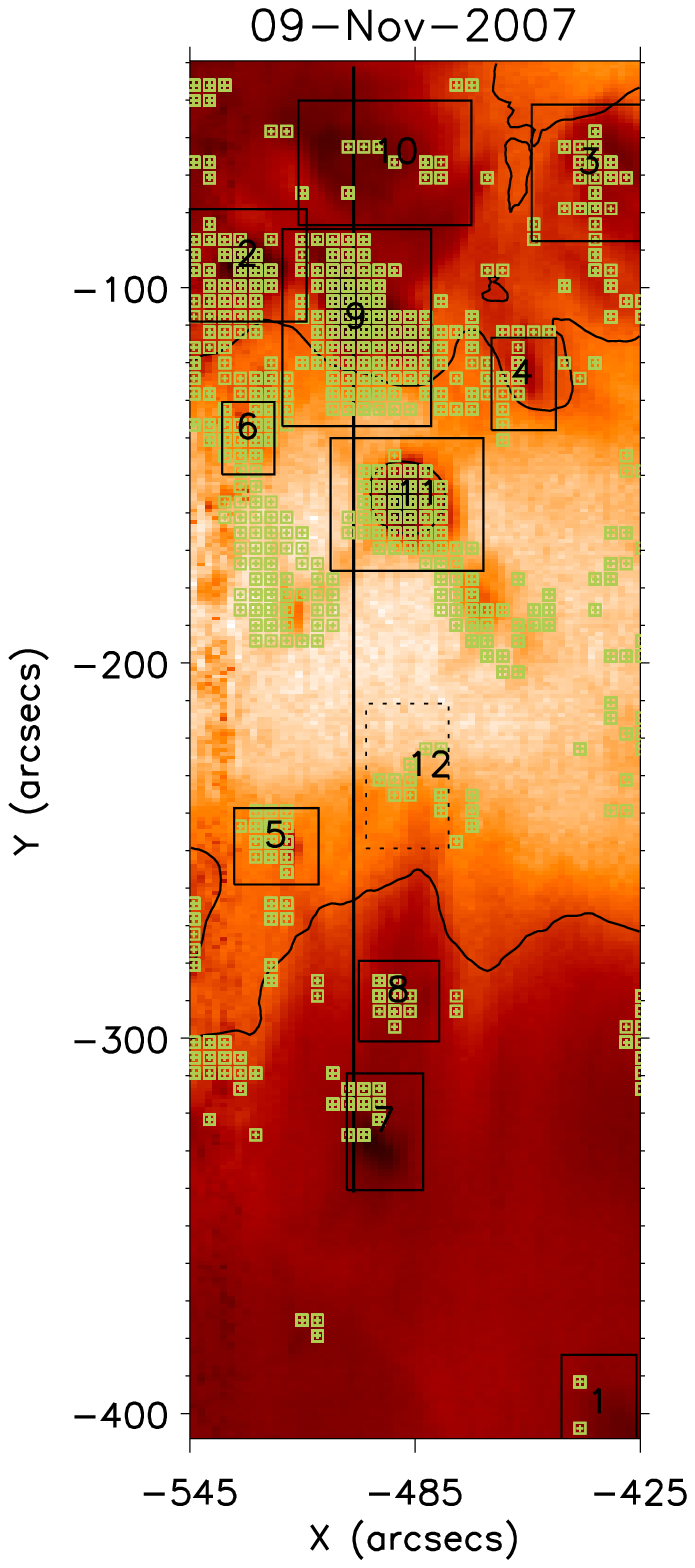}
\includegraphics{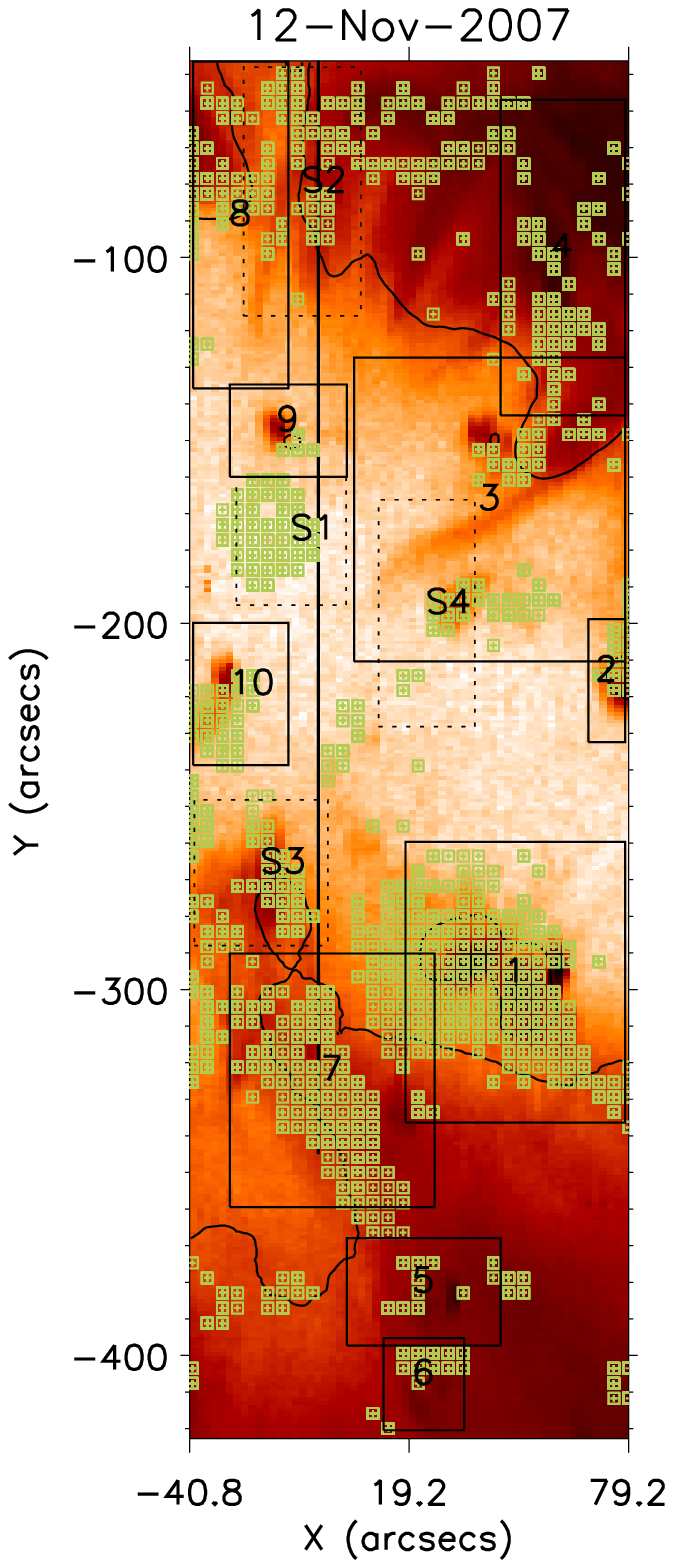}
\includegraphics{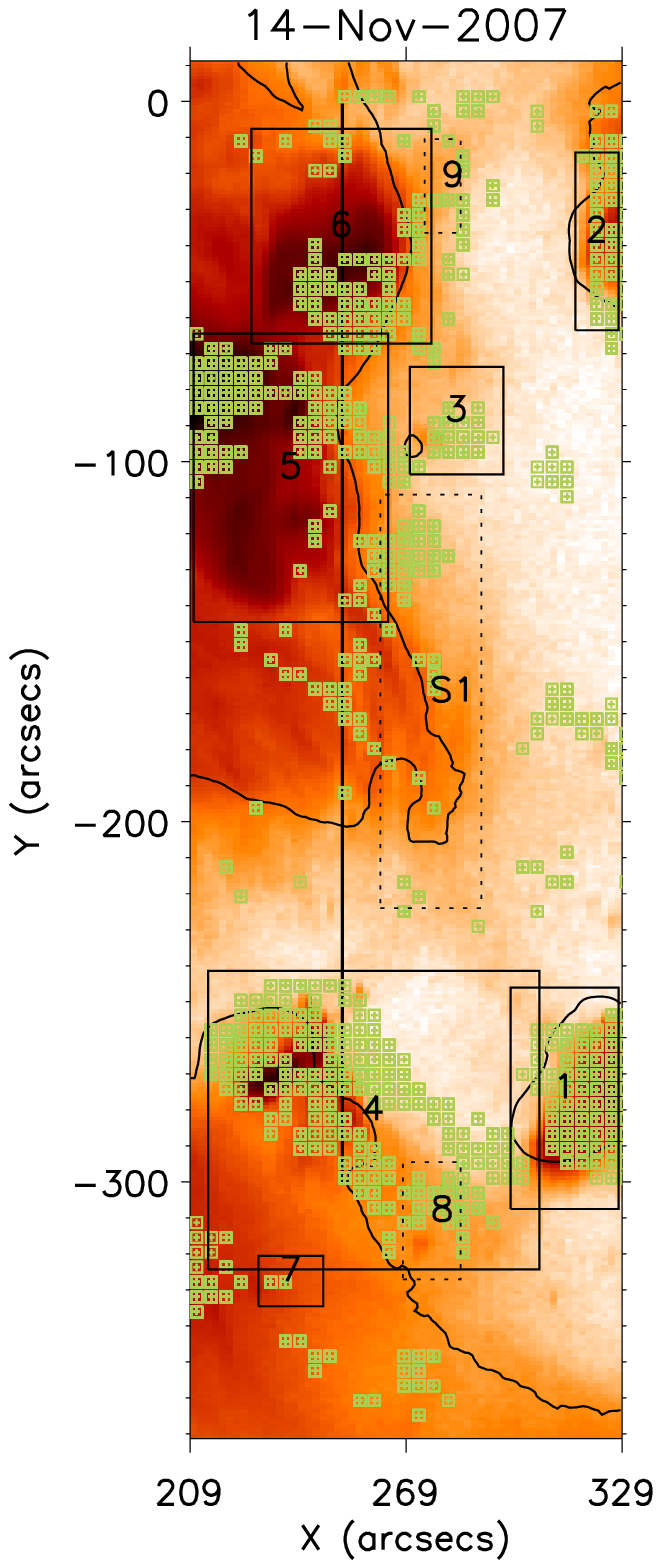}
\includegraphics{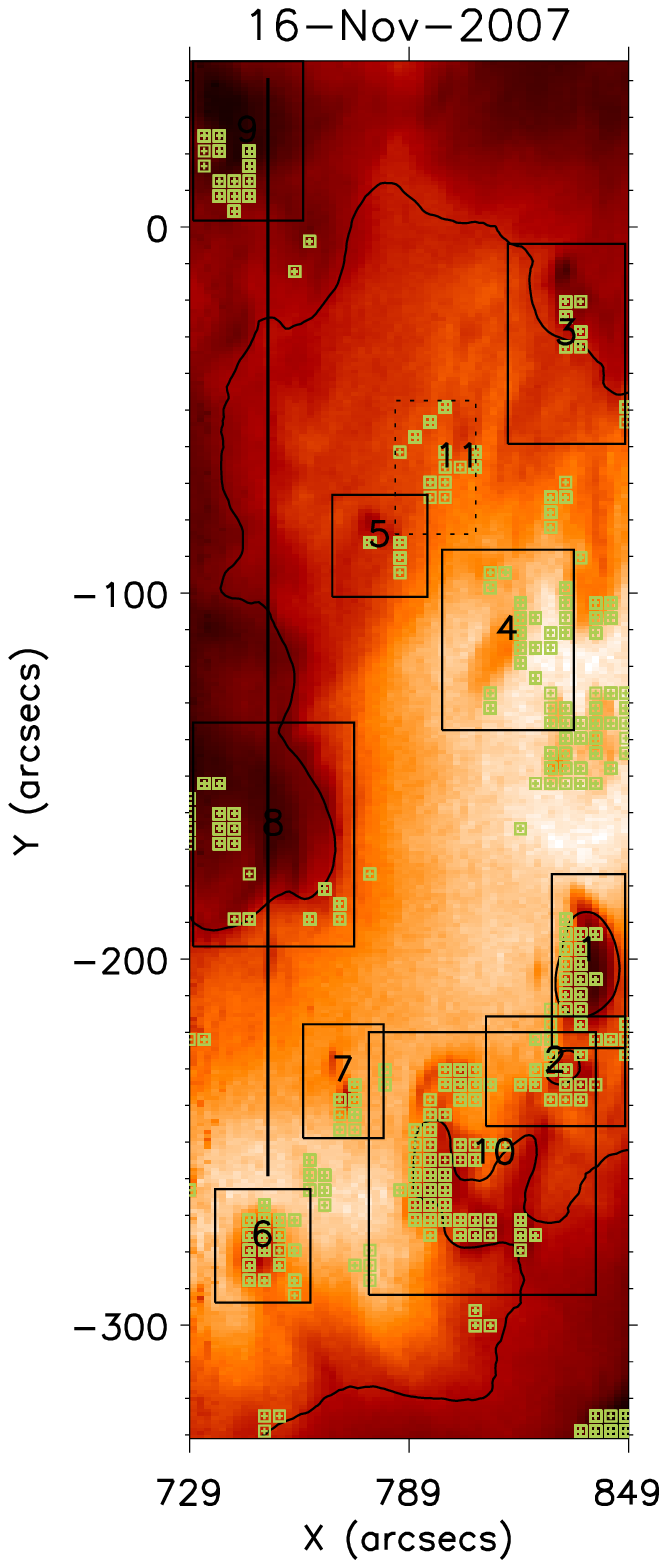}
   \caption{The EIS large rasters with the identified events over-plotted with solid line boxes. The dashed 
   line boxes indicate the events identified in the small rasters. The overplotted symbols show the pixels 
   identified with a brightening enhancement in the XRT data from paper~II.  }
   \label{fig3}
\end{figure*}

%%%%%%%%%%%%%%%%%%%%%%%%%%%%%%%%%%%%%%

\begin{table*}
\centering
%\begin{center}
\caption{Plasma parameters of the brightening events observed in the quiet Sun.``Later'' indicates that the outflow was seen after the EIS rastering was over and was only registered in XRT.}
\begin{tabular}{c c c c c c c}
\hline\hline
Date and &\multicolumn{2}{c}{Doppler shifts (km s$^{-1}$)}&Density& Line&Flow&Occurring rate\\
Event number& Blue& Red&Log$_{10}{N_e}$ (cm$^{-3}$)&at T$_{max}$&Outflow&\&Appearance\\
\hline
09/11/07&&&&&&\\
1&$-$5&--&8.61&Fe~{\sc xiii}&Y$^{later}$&$>$2; unresolved brightening\\
7&--&2.5&9.01&Fe~{\sc xv}&N&$>$1, BP with evolving loops\\
8&$-$3&3.5&8.58&Fe~{\sc xiii}&N&$>$1; unresolved brightening\\
10&$-$3&1.5&8.84&Fe~{\sc xv}&N&$>$2; brightening in QS loops\\
\hline
12/11/07&&&&&\\
4&$-$2&2&8.92&Fe~{\sc xiii}&N& numerous identified events in a large BP (small AR)\\
\hline
14/11/07&&&&&&\\
7&$-$2&--&8.65&Fe~{\sc xiii}&N&$>$2; point-like brightening\\
%10??????
\hline
16/11/07&&&&&&\\
9&$-$6&6&8.76&Fe~{\sc xiii}&Y&$>2$; large BP\\
\hline
10/01/09&&&&&&\\
1&$-$19/$-$12&80&9.38&Fe~{\sc xv}&N&1; brightenings in a loop\\
2&$-$2&--&8.78&Fe~{\sc xiii}&N&2; brightenings in a compact BP\\
3&$-$8/$-$5&7&8.70&Fe~{\sc xiii}&Y&$>$2; brightenings in a  small-scale loop\\
4&$-$2&5&8.85&Fe~{\sc xiii}&N&1; brightenings in part of a larger loop structure \\
\hline
13/01/09&&&&&\\
1&$-$3&--&9.25&Fe~{\sc xv}&Y&5; compact long-lived BP\\
2&$-$3&--&8.90&Fe~{\sc xiii}&Y&1; brightenings in a small-scale loop\\
3&$-$2&--&8.73&Fe~{\sc xiii}&N&1; brightenings in a small-scale loop \\
4&$-$6&8&9.08&Fe~{\sc xiii}&N&$>$2; small BP\\
\hline
\end{tabular}
\label{tb5}
\end{table*}
  
\section{Data analysis, results and discussion}

\begin{figure}[!htp]
\vspace{9.5cm}
  \centering
    \includegraphics{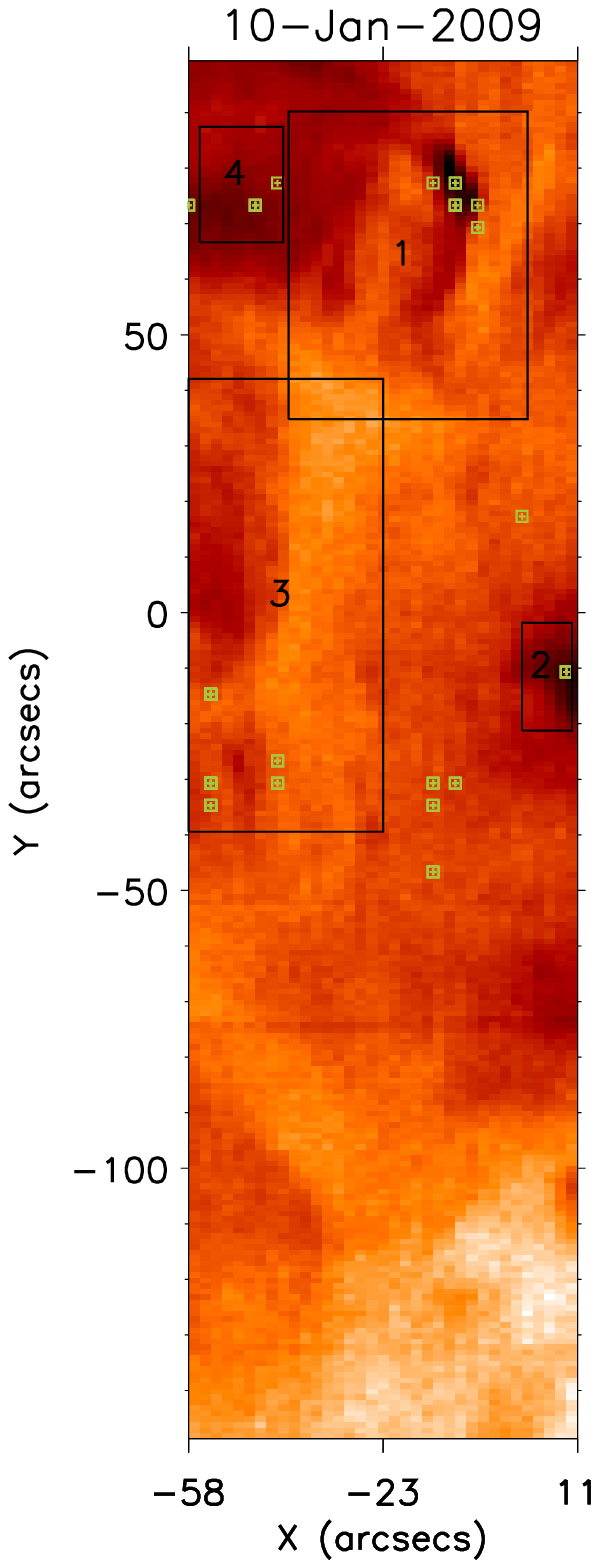}
\includegraphics{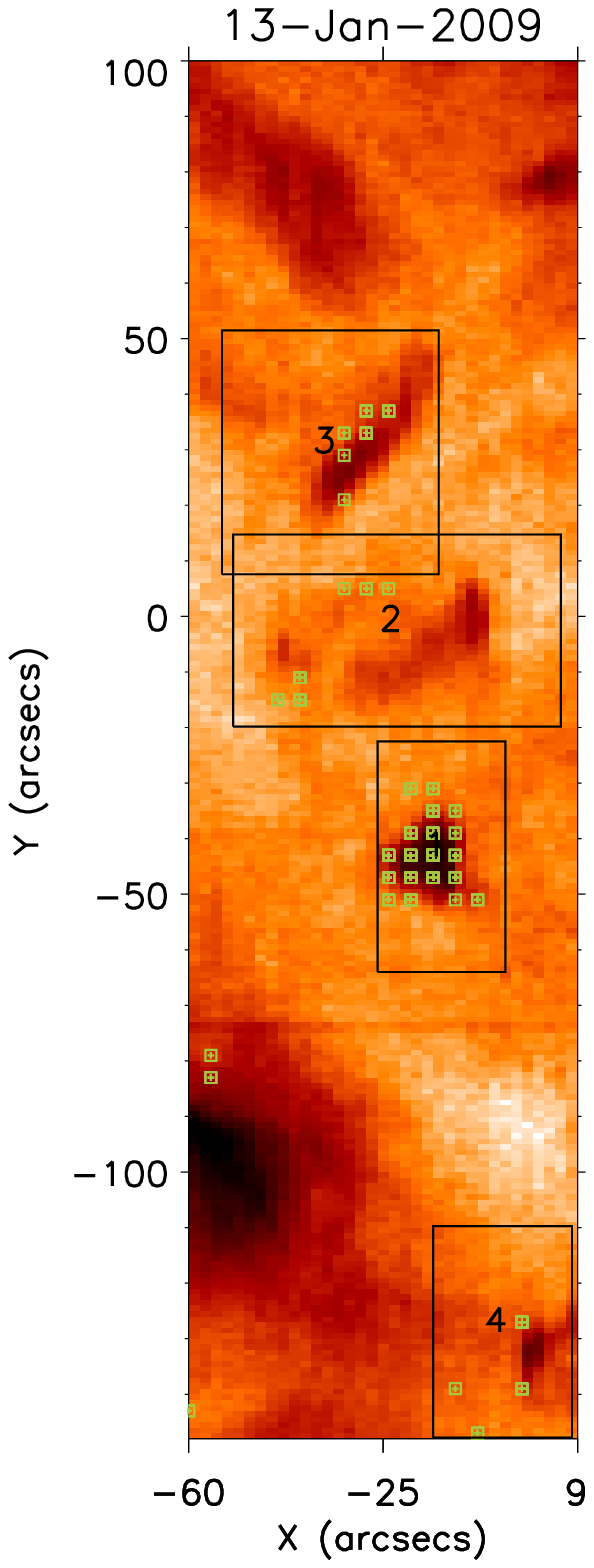}
   \caption{The EIS rasters on 2009 January 10 (left) and 13 (right). The overplotted symbols show the position 
   of the identified brightenings.}
      \label{fig4}
\end{figure}

The first step of our study was to align the EIS rasters and SUMER sit-and-stare data
with the XRT images analysed in paper~II  (see Fig.~\ref{fig1} and  Fig.~\ref{fig2}). Next, we located all brightenings 
identified in the XRT images (paper~II) on the EIS raster images. 
The grouping of brightening events was done as in paper~II, i.e.,  we inspected the
image sequence of each dataset with the identified brightenings
over-plotted at corresponding times (see Fig.~A\ref{fig17} and Fig.~A\ref{fig18}). Clusters of bright 
pixels identified next to each other and evolving at the same or at very close times were selected as 
one event. In Fig.~\ref{fig3} and Fig.~\ref{fig4}, the brightenings are overplotted on the CH and QS 
images, respectively.  Once each group of brightenings was unified into an individual event, each such 
event was given a number and it was marked as a quiet Sun,  a CHB or a CH event. A coronal hole boundary 
is defined as the region of $\pm$15\arcsec on both sides of the contour line outlining the coronal hole 
boundary as determined on the XRT images in paper~II. Part of the events registered in November 2007 
were also recorded by the SUMER spectrometer in the spectral lines shown in Table~\ref{tb2}. 
 We produced image sequences of all data taken at 
similar times in order to obtain a visual picture  of the event evolution. The combination of spectroscopic 
and imager data presented in this manner is the best existing way to obtain a full description of an 
observed phenomenon using the presently existing solar instrumentation. Examples of this kind of  
`movies' are given in the online material ~(see Fig.~A\ref{fig17} and Fig.~A\ref{fig18}).\footnote{Movies for each dataset in 2007 and 2009 can be found at \url{http://star.arm.ac.uk/highlights/2012/599-media/}} We then derived the physical plasma parameters 
of each event, i.e. their  structural evolution in time, Doppler shifts,  temperatures and densities. Because of the 
way slit spectrometers operate as well as the random occurrence of transient brightenings, we could  
`catch'  the physical parameter evolution in various parts of these events and at various stages of the event's  
development. Unfortunately, the event best observed by all instruments remains the one discussed in 
detail  by \citet{madj2010}. 

In Figs.~\ref{fig1} and \ref{fig2} we present the XRT images studied in paper~II which were used to automatically 
identify transient brightenings.  From the inspection of the composed image animations, we were able to visually 
determine the spectroscopic signature  of  each brightening. The analysis is not trivial and very time 
consuming because it requires the consideration of each type of observation, i.e. a rastering and a sit-and-stare, 
the exposure duration, contaminations from line-of-sight contributions and last but not least various instrumental 
effects which are specific for each instrument. The number of QS events  which were found in the November 
2007 data is  small in comparison  with the number of CH events because of the limited FOV of the instruments.  
In order to increase the number of analysed QS events, we also used the two QS  datasets taken on 2009 
January  10 and 13. The EIS observations for this run were done with a different program, i.e. only larger rasters  
were taken. Nevertheless, these data provide  the same quality information as the November 2007 run. In 
Figs.~\ref{fig3} and \ref{fig4}, the EIS FOVs (only the big rasters for November 2007) are shown for all observations.  
The brightening pixels as identified through an automatic procedure (paper~II) in XRT Al$\-$poly images are 
overplotted with symbols. The solid and dashed (only in Fig.~\ref{fig3}) line boxes correspond to the 
identification during the big  and small rasters, respectively. The coronal hole boundaries are defined as 
to match the boundaries considered in paper~II. All physical parameters are given in Tables~\ref{tb5}, \ref{tb6} 
and \ref{tb7} and are discussed in detail in the next sections.

\subsection{Quiet Sun brightenings}

\subsubsection{Structure and dynamics}

The quiet Sun brightenings (15 in total) all evolve starting with a brightening in an area of a few pixels 
(microflaring).  The evolution proceeds in several patterns which seem to depend on the size of the phenomenon: 
(1) The initial brightening can be seen  spreading along a loop system or just a single loop (observed at the EIS 
and XRT spatial resolution);  (2) It may evolve from  a point-like brightening ($\sim$3--4 pixels)  which  may enlarge 
slightly and then quickly (1--2 min) fade away; (3) Microflaring occurs in pre-exiting BPs which triggers a 
brightening increase over the whole or large parts of the pre-exiting BP. In the latter case, loops are often 
seen to expand from the BP and then shrink back to their original size. In many cases no feature is present at X-ray 
temperatures prior to the transient brightening.  That clearly indicates that after energy deposition(s), the plasma 
was heated to X-ray temperatures and/or then  ejected along the magnetic field-lines of pre-existing or newly formed loops.  

\begin{figure}[!h]
    \includegraphics[scale=0.48]{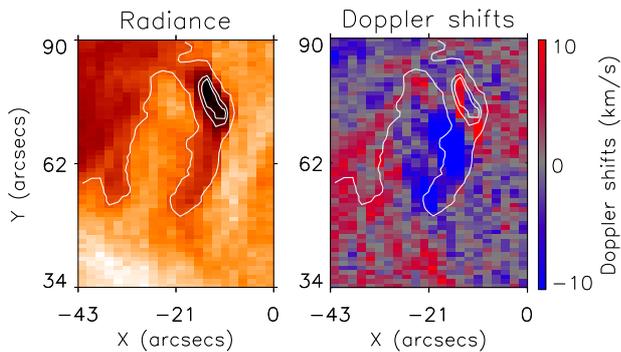}
 \caption{Event No. 1 on January 10. The raster was taken in the Fe~{\sc xii} 195~\AA\ line between 12:35~UT and 12:59~UT. The left and right panels represent the Fe\,{\sc xii}\,195\AA\, intensity and Doppler shift images, respectively. The intensity image is plotted in negative logarithmic scale.}
   \label{fig5}
\end{figure}

\begin{figure}[!h]
    \hspace*{-0.5cm}
\includegraphics[scale=0.55]{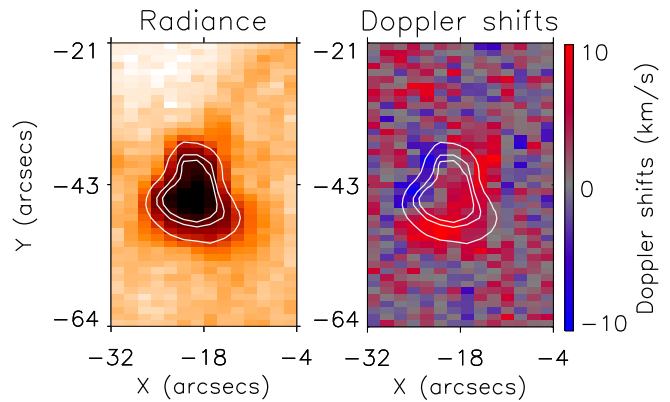}
   \caption{Event No. 1 on January 13. The raster was taken in the Fe~{\sc xii} 195~\AA\ line between 15:01~UT and 15:12~UT.}
   \label{fig6}
\end{figure}

\begin{figure}[!h]
    \hspace*{-0.3cm}
\includegraphics[scale=0.49]{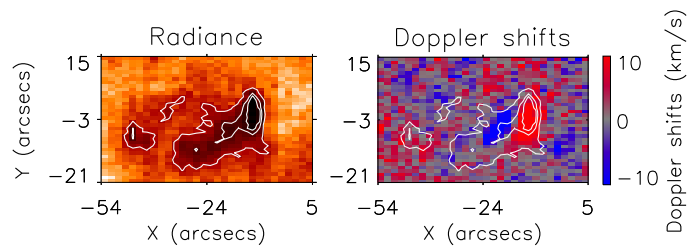}
   \caption{Event No. 2 on January 13. The raster was taken in the Fe~{\sc xii} 195~\AA\ line between 14:35~UT and 14:55~UT.}
   \label{fig7}
\end{figure}

To investigate the plasma properties and dynamics of the transient brightenings we analysed EIS Doppler shift 
maps and intensity images in spectral lines covering a large temperature range (see Tables~\ref{tb1} and \ref{tb2}) 
together with co-temporally taken  X-ray images. This way of analysing transient events permits to 
separate a brightening  expansion from a plasma flow.  The online material provides some of the observational material 
used in the present study (Fig.~A\ref{fig17} and Fig.~A\ref{fig18}). In addition, we present three events in  Fig.~\ref{fig5}, \ref{fig6} and \ref{fig7} which are typical examples of  QS transient phenomena. One should bear in mind that the Doppler velocity images were produced from a single Gauss fit and, therefore, do not reflect the exact  pattern of the plasma flows. 
These Doppler images, however, give a good indication of the dominant flows during the rastering of the events. 
 
The event in Fig.~\ref{fig5} started with a brightening in a single pixel as seen in the XRT image at 12:20~UT (Fig.~A\ref{fig17}). The brightening spread over more pixels until  a loop structure is clearly visible in the XRT image  taken 6 min later 
(Fig.~\ref{fig5}). During the EIS scan starting at 12:35~UT, the feature shows clear red- and blue-shift patterns.  
The red-shift coincides with the energy deposition site while the blue-shift appears along a loop structure originating 
from the energy deposition site.  Several more brightenings re-occurred in the same structure but they  were not 
identified as brightenings in the XRT data in paper~II because they did not meet the selection criteria adopted there.  
The flows remain trapped in closed loops.
  
 %%%%%%%%%%%%%%% Fig 4 %%%%%%%%%%%%%%%%%%%
%%%%%%%%%%%%%%%%%%%%%%%%%%%%%%%%%%%%%%

%%%%%%%%%%%%%%% Fig 5 %%%%%%%%%%%%%%%%%%%

In Fig.~\ref{fig6} we show a feature which is a typical example of  a coronal bright point (BP) with a diameter of 
less than 20\arcsec. The event was scanned by EIS during several rasters and the first feature to notice is that 
the general blue-red Doppler-shift pattern of this phenomenon remains very similar during the entire lifetime of 
the BP, i.e. half of the brightening shows a blue-shifted emission and the other half a red-shifted emission. 
These Doppler-shifts are well known to indicate up-flows and down-flows of both sides of loops. However, as 
discussed in \citet{2012ApJ...744...14Y} blue/red-shift Doppler patterns can often occur due to strong 
intensity gradients. It is caused by an asymmetric point 
spread function (PSF) which is discussed in detail in   \citet{2012ApJ...744...14Y}.  In order to avoid a 
mis-interpretation we cross-checked each Doppler shift with imaging information from XRT. In the cases where
the image sequence did not confirm  a plasma proper motion while Doppler shift was measured,  this Doppler 
shift was not considered as real but rather instrumental (PSF). The BP had several episodes of microflaring 
which resulted in the BP becoming larger with small loops rising and than shrinking  back to their original size.  
None of the microflaring resulted in  the formation of an EUV/X-ray jet.

The event in Fig.~\ref{fig7} is a very clear example of a brightening followed by an expanding loop. This event, 
however, suffered by a gap of 9 min in the XRT observations. Nevertheless, we clearly see that a brightening  
in a region of a few pixels is followed by  a loop expansion originating from this region. Similar examples 
will be studied best in the future using AIA (The Atmospheric Imaging Assembly) data were rising and shrinking of loops can be separated from 
cooling and heating thanks to the unprecedented cadence of the instrument as well as its numerous passbands 
coverage. The blue- and red-shift patterns in Fig.~\ref{fig7} are difficult to interpret but they seem to be 
emitted from the rising and than contracting loop as seen in the XRT image sequence. Note that an exposure 
time of 40~s was used here which is relatively long with respect to the dynamics of the observed phenomena.

The plasma properties of the events found in the quiet Sun are listed in Table~\ref{tb5}. The Doppler shifts
are measured as an average over the  blue- or red-shift pattern associated with the event. The electron 
density was calculated  over only the bright pixels in order to obtain sufficient signal-to-noise. Each spectral 
line was investigated for a response in the studied event and thus the maximum temperature of the event was defined. The temperatures and densities of the quiet Sun events are discussed in the next Section.

In general the QS phenomena evolve dynamically, i.e. small-scale structural changes are seen at any time.  
Despite this `activity', however, none of the QS events  has been identified as a jet like event (see the 
definition in Sect.~\ref{intro}). The average Doppler shifts of the QS events are a blue-shift of  $-$3.6~\kms\ 
and a red-shift of 4.4~\kms\ derived from a single Gauss fit. One should bear in mind that these are not absolute 
values (see Sect.~2.1 for more details) but rather give an indication of the velocity range of QS brightening 
events. From 15 brightenings, five show flows/outflows, i.e. plasma is seen to move away from the initial 
source brightening. In all cases, however, the ejected plasma remains trapped in loops. In one event 
the outflow was observed in the XRT images after the EIS rastering of this region was over. The events which 
are not seen to produce a flow or outflow in the XRT images appear as a compact brightening  which is maybe 
due to their intricate structure or simply not sufficient instrumental resolution. There exist the probability 
that in some cases the flows/outflows had far too low emission and, therefore, could not be separated from the 
background emission. A future multichannel AIA study should confirm this possibility. All events occurred 
several times as visually estimated  from their XRT lightcurves (for more details on this see paper~II). We do 
not give a precise number of the occurrence rate  as that will require a definition of a time and intensity  
threshold which is not a subject of this study and has been done in detail in paper~II. This information though 
gives an indication of the dynamics of the events studied here.

%%%%%%%%%%%%%%% Fig 6 %%%%%%%%%%%%%%%%%%%

%%%%%%%%%%%%%%% Fig 7 %%%%%%%%%%%%%%%%%%%

\subsubsection{Temperatures and electron densities}
\label{qstnne}
The temperature of the quiet Sun events is defined by the maximum formation  temperature of the spectral lines in 
which these events were registered (Table~\ref{tb5}).  All observed spectral lines were investigated for a response in each event.
From 15 events, 4 are seen in Fe~{\sc xv} (logT$_{max}$$\approx$6.3~K) 
and  11 in Fe~{\sc xiii}  (logT$_{max}$$\approx$6.2~K). None of the events showed  signature in spectral lines with formation temperatures higher than the formation temperature of Fe~{\sc xv}. These temperatures refer in most cases to the energy deposition 
 site (the pixels with the highest intensity) which always has the highest temperature. 

The electron density was obtained from the spectral line ratio of Fe~{\sc xii} 186.88~\AA\ to 195.12~\AA\ (for more 
details see \citet{madj2010}) and describes plasmas at temperatures logT$_{max}$$\approx$6.1~K  which is the maximum 
formation temperature of Fe~{\sc xii}. The quiet Sun events have an average electron density of 
log$_{10}{N_e}$ $\approx$ 8.87~cm$^{-3}$. In only four cases, the electron density is above 
log$_{10}{N_e}$ $\approx$ 9.0~cm$^{-3}$ (Table~\ref{tb5}: event 7 on Nov 9, event 1 on Jan 10 and events 1 and 4 on 
Jan 13). The event 7 is a BP clearly present in X-rays. The measurement of the electron density for event 1 on Jan 
10 were taken during the peak of the intensity as defined from the X-rays light-curve. Both events 1 and 4 on Jan 13 are  BPs.

  \subsection{CH and CHB brightenings}
 
 \subsubsection{Structure and dynamics}

The CH events, 35 in total, were concentrated at the coronal hole boundaries and inside coronal holes where 
 pre-existing or newly emerging loop structures known as coronal bright points are found. From 35 events 23, 
i.e., 65$\%$ were seen to produce outflows which in most cases represent  jet-like phenomena called EUV/X-ray 
jets as observed in  XRT images. Several features were found to be common for all observed events. All were 
associated with  coronal bright points either pre-existing or newly emerging at X-ray temperatures. They evolve 
dynamically and produce numerous outflows (jet-like phenomena) until the full disappearance of the BP. Multiple 
energy deposition during a single event is a common feature. After each jet, the BP becomes smaller or even 
invisible at X-ray temperatures. In several cases  BPs are not visible at XRT temperatures prior to the transient 
event  because they are formed of cooler lower lying loops. We refer to these BPs as newly emerging. These 
BPs would appear as a small-scale (only a few pixels) brightening in the XRT images, and a few minutes or 
even seconds prior the occurrence of the X-ray jet . After the jet, the BP will again become invisible at X-ray temperatures.  
Several events of this type can be followed in the online material for the four days of coronal hole observations (see
Fig.~A\ref{fig17} and A\ref{fig18}). All  events follow the same evolutionary scenario, i.e.,  a sudden 
brightening (microflaring) in a BP is followed by plasma ejections as in the case described in \cite{madj2010}.  
 
  %%%%%%%%%%%%%%% Fig 8 %%%%%%%%%%%%%%%%%%%

 \begin{figure}[!ht]
\includegraphics[scale=0.45]{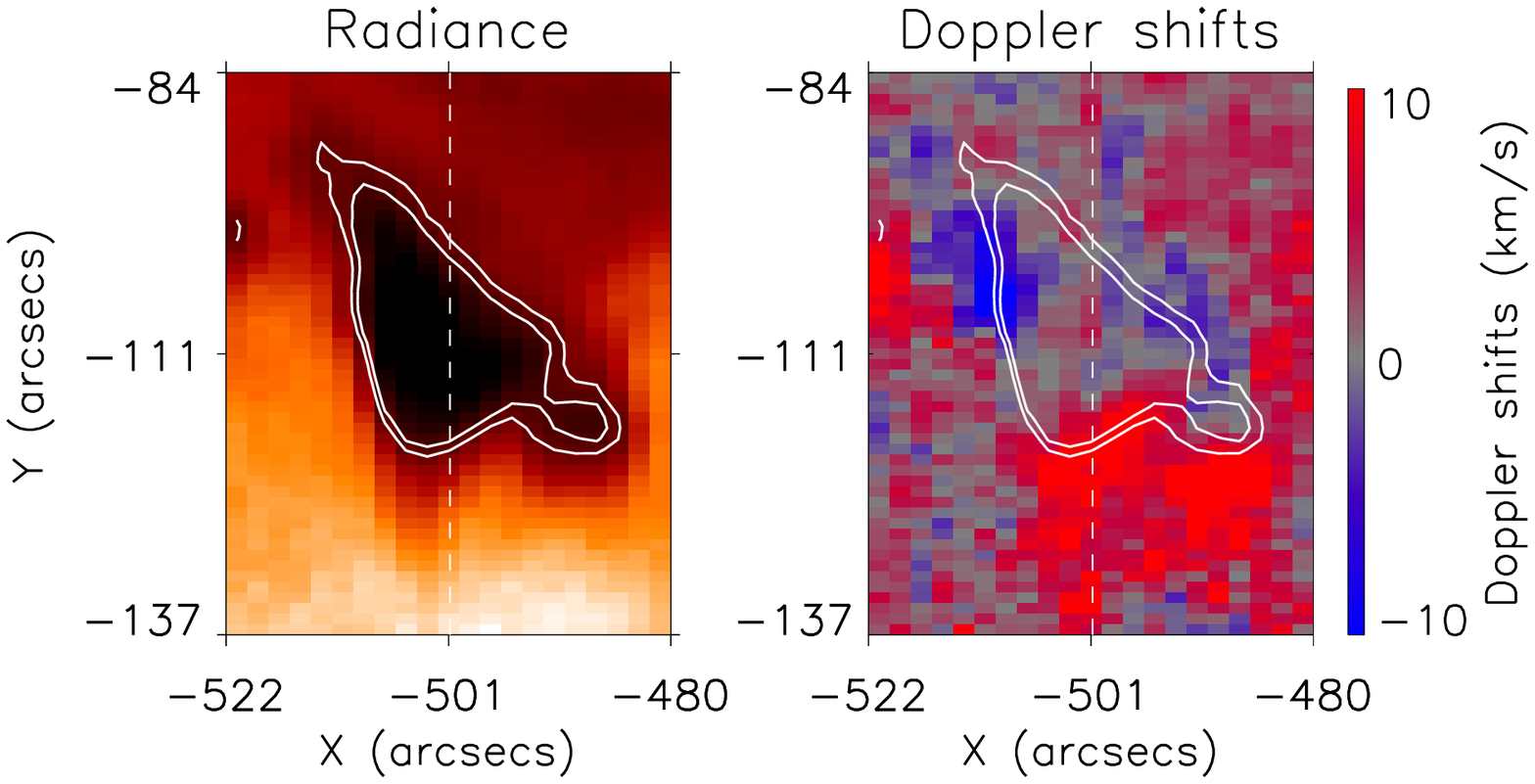}\\%CHB
\includegraphics[scale=0.45]{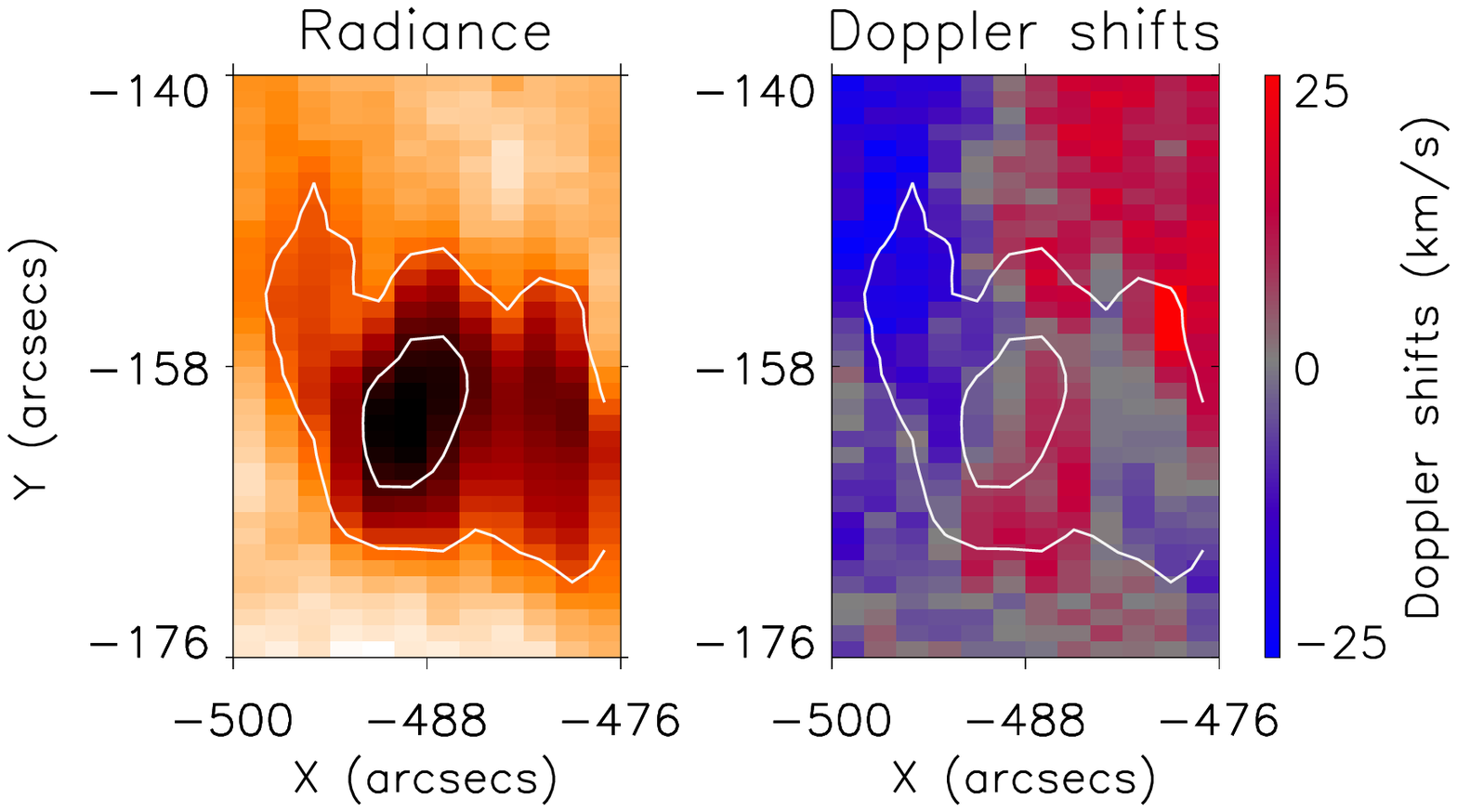}%CH
\caption{Two examples, event 9 (top) and 11 (bottom) respectively of brightening events  observed on November 
9. The left and right panels represent the Fe~{\sc xii}~195~\AA\ intensity  and Doppler shift images, respectively. 
The event 9 raster is taken between 07:03~UT and 07:24~UT and the raster for event 11 is from 11:47~UT  
to 11:59~UT (see the online material to follow its full evolution). Event 9 is at the coronal hole boundary  
while event 11 is inside the coronal hole. White dashed lines indicate the SUMER slit sit-and-stare position. 
The radiance images are plotted in negative logarithmic scale.}
\label{fig8}
\end{figure}

%%%%%%%%%%%%%%% Fig 9 %%%%%%%%%%%%%%%%%%%
\begin{figure}[!ht]
\includegraphics[scale=0.45]{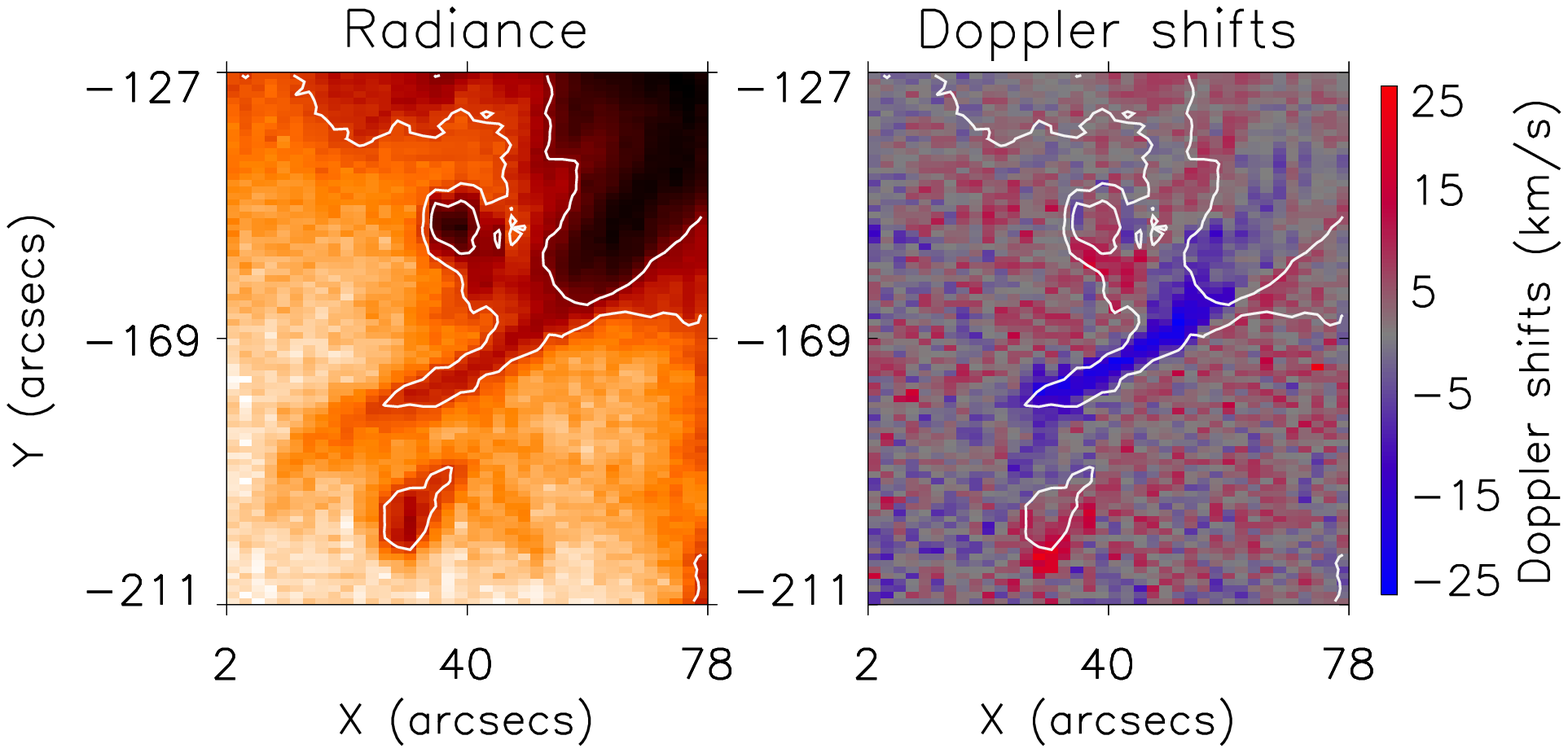}\\ %CHB
\includegraphics[scale=0.45]{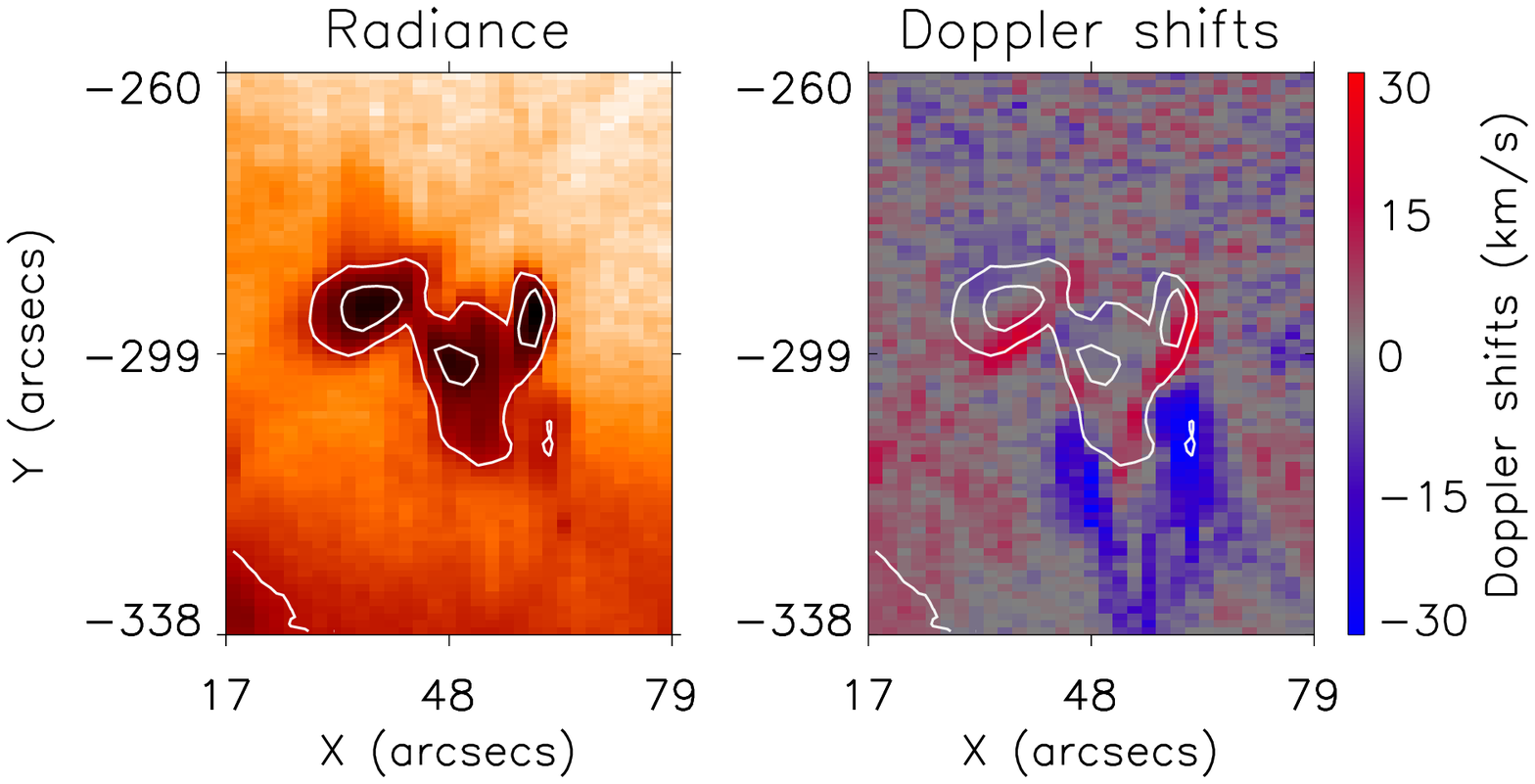} %CH
     \caption{Two examples, brightening event 3 at coronal hole boundaties~(top) and 1 in coronal hole~(bottom), observed on November 12. The left and right panels represent the Fe~{\sc xii}~195~\AA\ intensity  and Doppler shift images, respectively. Event 3 was 
     scanned between 01:17~UT and 01:54~UT while event 1 between 01:17~UT and 01:49~UT during the big raster 
     observations. The top event happens at the coronal hole boundary while the bottom inside the coronal hole.}
   \label{fig9}
\end{figure}

 \begin{table*}%{sidewaystable}
%\begin{center}
\centering
\caption{As table \ref{tb5}, presenting the plasma properties of the  events observed inside coronal holes and at coronal hole boundaries.}
\begin{tabular}{c c c c c c c c}
\hline\hline
Date and &\multicolumn{2}{c}{Doppler shifts (km s$^{-1}$)}&Density&Line&Flow/& Occurrance rate,\\
Event number&Blue&Red&Log$_{10}{N_e}$ (cm$^{-3})$&at T$_{max}$&Outflow& Appearance, Phase\\
\hline
09/11/07&&&&&&&\\
2&$-$140/$-$8&4.8&9.23&Fe~{\sc xv}&Y&$>2$; compact BP; post brightening phase\\
3&$-$5&4&8.98&Fe~{\sc xiii}&N&$>2$; pre-activity phase, only brightening, no jet yet; BP\\
4&$-$2&2&8.96&Fe~{\sc xiii}&Y$^{later}$&2; point-like brightening, quiet phase\\
5&--&3/77&9.32&Fe~{\sc xiii}&Y&2,1 large; point-like brightening\\
6&$-$5&3/180&9.21&Fe~{\sc xiii}&Y&$>1$, 1 large, flow in a small-scale loop\\
9&$-$199/$-$3&15&9.24&Fe~{\sc xv}&Y&$>3$, 2 large; large BP\\
11&$-$160/$-$9&12&9.48&Fe~{\sc xv}&Y&$>3$; long-lived BP\\
12&$-$160/$-$8&2&8.60&Fe~{\sc xiii}&Y&1; jet from point-like brightening\\
\hline
12/11/07&&&&&&&\\
1&$-$170/$-$20&15/76&9.37&Fe~{\sc xv}&Y& $>$3; large jets, slit crosses the whole event \\
2&--&2.6&9.26&Fe~{\sc xiii}&N&$>$3 later; quiet phase\\
3&$-$170&7&8.76&Fe~{\sc xiii}&Y& 1 large but faint, slit crosses the jet \\
5&--&8&8.85&Fe~{\sc xv}&N& $>$ 2; point-like brightening\\
6&$-$10/$-$150/&--&8.66&Fe~{\sc xv}&Y&2; slit crosses only the outflow\\
7&$-$140/$-$20&--&8.74&Fe~{\sc xiii}&Y&1 large; slit crosses only the jet\\
8&--&22&8.91&Fe~{\sc xiii}&Y$^{later}$& 4; quiet phase, small brightenings\\
9&--&10&9.56&Fe~{\sc xiii}&N&1; point-like brightening\\
10&$-$120&6&9.34&Fe~{\sc xiii}&Y$^{later}$&$>$2; point-like brigntening\\
\hline
14/11/07&&&&&&\\
1&--&5&9.41&Fe~{\sc xv}&Y$^{later}$&$>2$ large jets, pre-activation phase\\
2&$-$2&2&9.09&Fe~{\sc xiii}&Y$^{later}$&$>1$, evolving loops of a faint BP\\
3&$-$87&182&8.97&Fe~{\sc xiii}&Y$^{small}$&$>3$; small BP\\
4&$-$270&15&10.11&Fe~{\sc xxiii}&Y&$>3$, 2 observed jets\\
5&--&2&9.25&Fe~{\sc xvi}&Y$^{later}$&$>3$, long-lived BP\\
6&--&2&9.14&Fe~{\sc xv}&Y$^{later}$&$>2$, long-lived BP\\
8&--&10&9.23&Fe~{\sc xiii}&N&$>2$; point-like brightening\\
9&--&13&9.21&Fe~{\sc xiii}&N&$>2$; appears as brigtenings in a small-scale loop \\
%13???????????
\hline
16/11/07&&&&&&\\
1&$-$5&6&9.15&Fe~{\sc xv}&Y$^{later}$&2; BP in a quiet phase\\
2&--&8&9.07&Fe~{\sc xiii}&Y$^{later}$&$>1$;BP in a quiet phase\\
3&$-$2&5&9.03&Fe~{\sc xiii}&N&$>2$; quiet BP\\
4&$-$3&3&8.89&Fe~{\sc xiii}&Y$^{later}$&1; faint loop brightenings\\
5&$-$2&2&8.55&Fe~{\sc xiii}&N&$>1$; point-like BP\\
6&--&2&8.98&Fe~{\sc xiii}&N&$>3$; small BP, quiet phase\\
7&--&12&9.03&Fe~{\sc xiii}&N&3; point-like brightening\\
8&$-$11&6&8.84&Fe~{\sc xv}&Y$^{later}$&1 large, small AR\\
10&$-$5&8&8.80&Fe~{\sc xv}&Y$^{small}$&$>3$; large BP in a quiet phase\\
11&$-$3&4&8.82&Fe~{\sc xiii}&N&$>2$; point-like BP \\
\hline
\end{tabular}
%\end{center}
\label{tb6}
\end{table*}
%Doppler shifts were measured from those pixels of each event which are better to apply triple Gaussian fits by using EIS observed Fe~{\sc xii} 195.12 \AA. Red-shifted and Blue-%shifted were determined from the residual of single Gaussian fit. Where the single Gaussian fits were applied by a constant $\sigma$ and line center.

In  order to obtain the Doppler shifts in high velocity events, we used a multi-Gaussian fit (double or triple) for EIS Fe~{\sc xii}~195.12~\AA\ and SUMER O~{\sc v}~629.77~\AA. The largest red- and blue-shifts correspond of the peak of the Gaussian fit in the corresponding wing of the line. The wings extended to even higher 
velocity but  the low signal does not permit a  Gaussian fit.  In most of the cases we were also able to fit the blend by Fe~{\sc xii}~195.18~\AA\ \citep{2009A&A...495..587Y}. The average Doppler shifts of the CH events are 
a blueshift of $-$160~\kms\ and red-shift of 17~\kms.  These large Doppler velocities clearly indicate the more 
dynamic nature of CH/CHB events with respect to the QS phenomena. In all cases, the high Doppler shifts were found 
only in  the jet-like events. This strongly suggests that the co-existence of open and closed magnetic fields 
is a necessary condition for plasmas to be accelerated to high velocities.

Interestingly the number of registered jets has decreased while the coronal hole was progressing through the 
visible solar disk towards the limb. One possible explanation is that the change of the line-of-sight is the 
cause. Another possibility is a decrease of ephemeral region emergence in the coronal hole with time but 
this has never been reported or studied as far as we are acquainted of.

%%%%%%%%%%%%%%% Fig 10 %%%%%%%%%%%%%%%%%%%
\begin{figure}[!ht]
\includegraphics[scale=0.45]{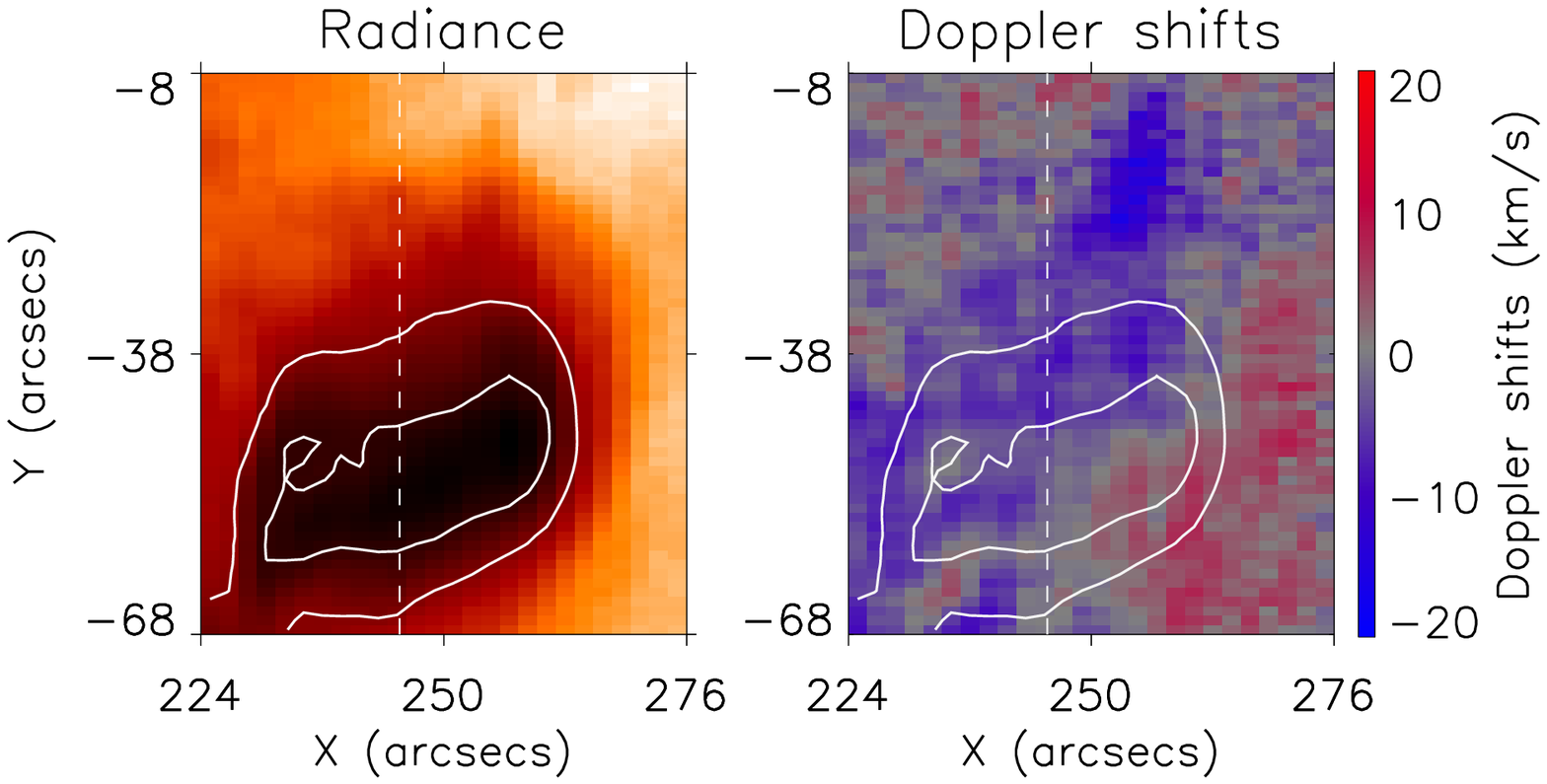}\\ %CHB ; old version is No. 15
\includegraphics[scale=0.45]{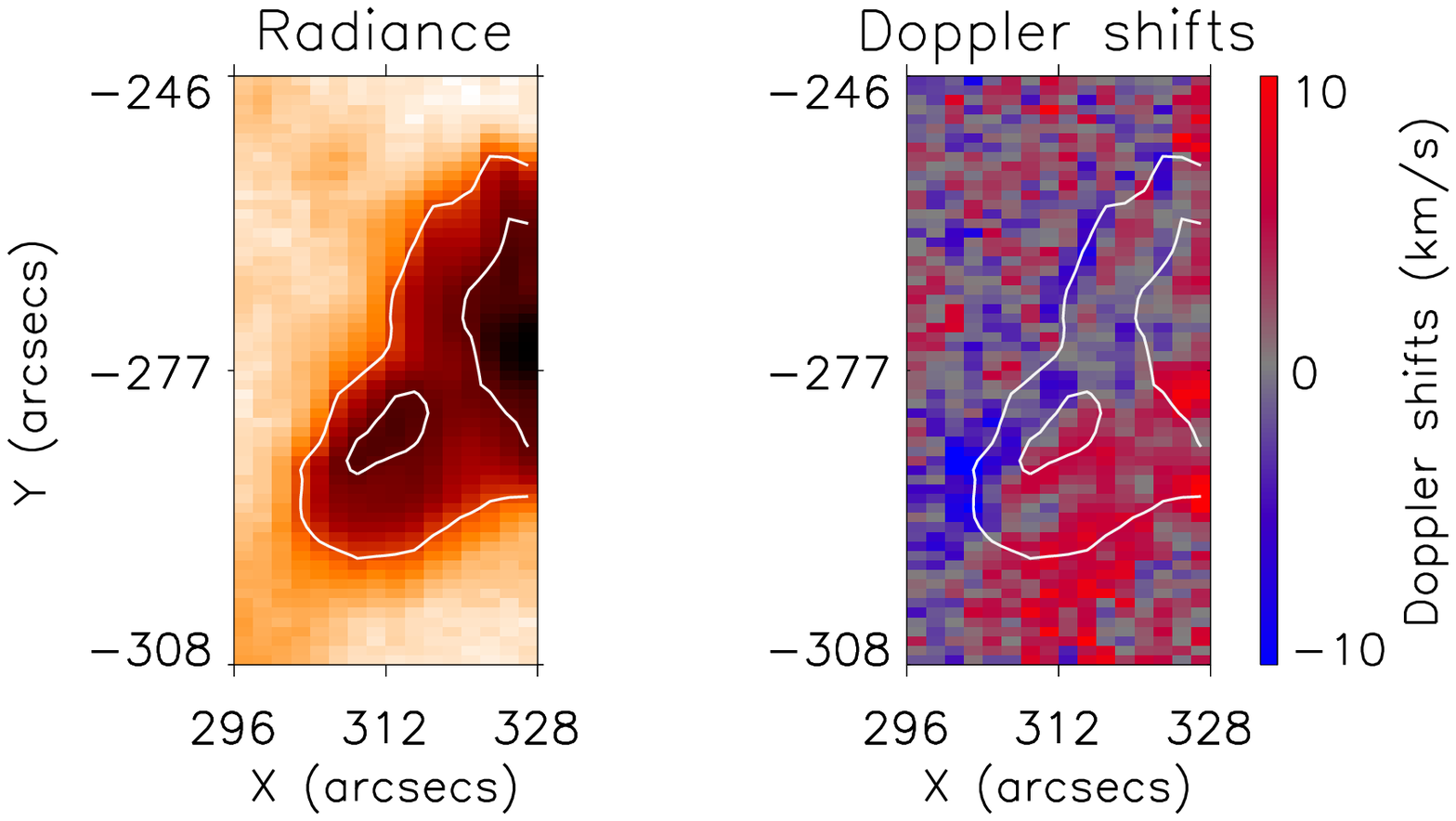}%CH4
\caption{Two examples of brightening events observed on November 14 (top: event 6 at a coronal hole boundary; bottom: event 1 inside the coronal hole). The left and right panels represent the Fe~{\sc xii}~195~\AA\ intensity and Doppler 
shift images, respectively.  Event 6 was observed between 00:48~UT and 01:15~UT while event 1 from 00:17~UT until 00:40~UT.}
   \label{fig10}
\end{figure}

%%%%%%%%%%%%%%% Fig 11 %%%%%%%%%%%%%%%%%%%
\begin{figure}[!ht]
\includegraphics[scale=0.45]{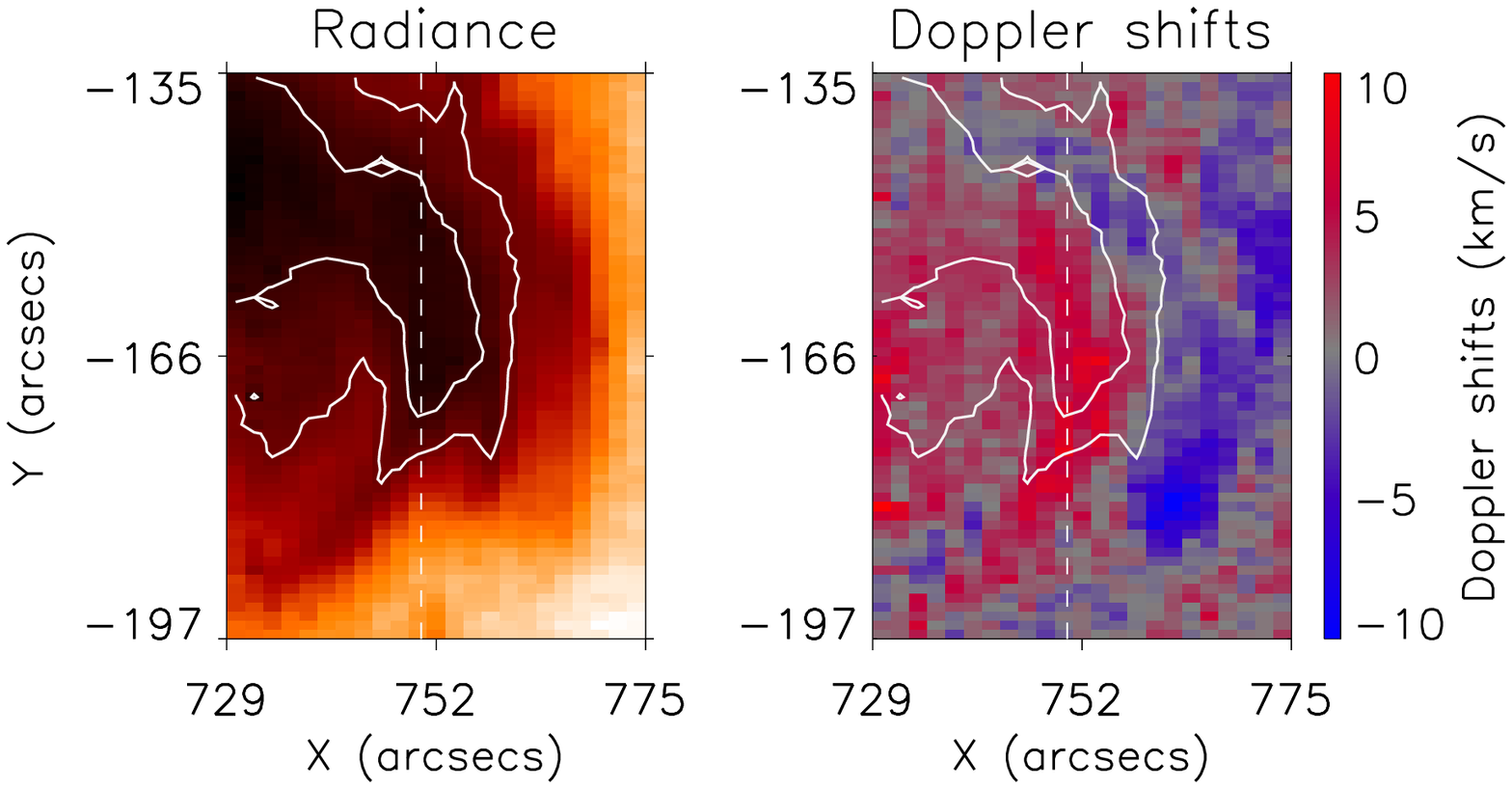}\\
\includegraphics[scale=0.45]{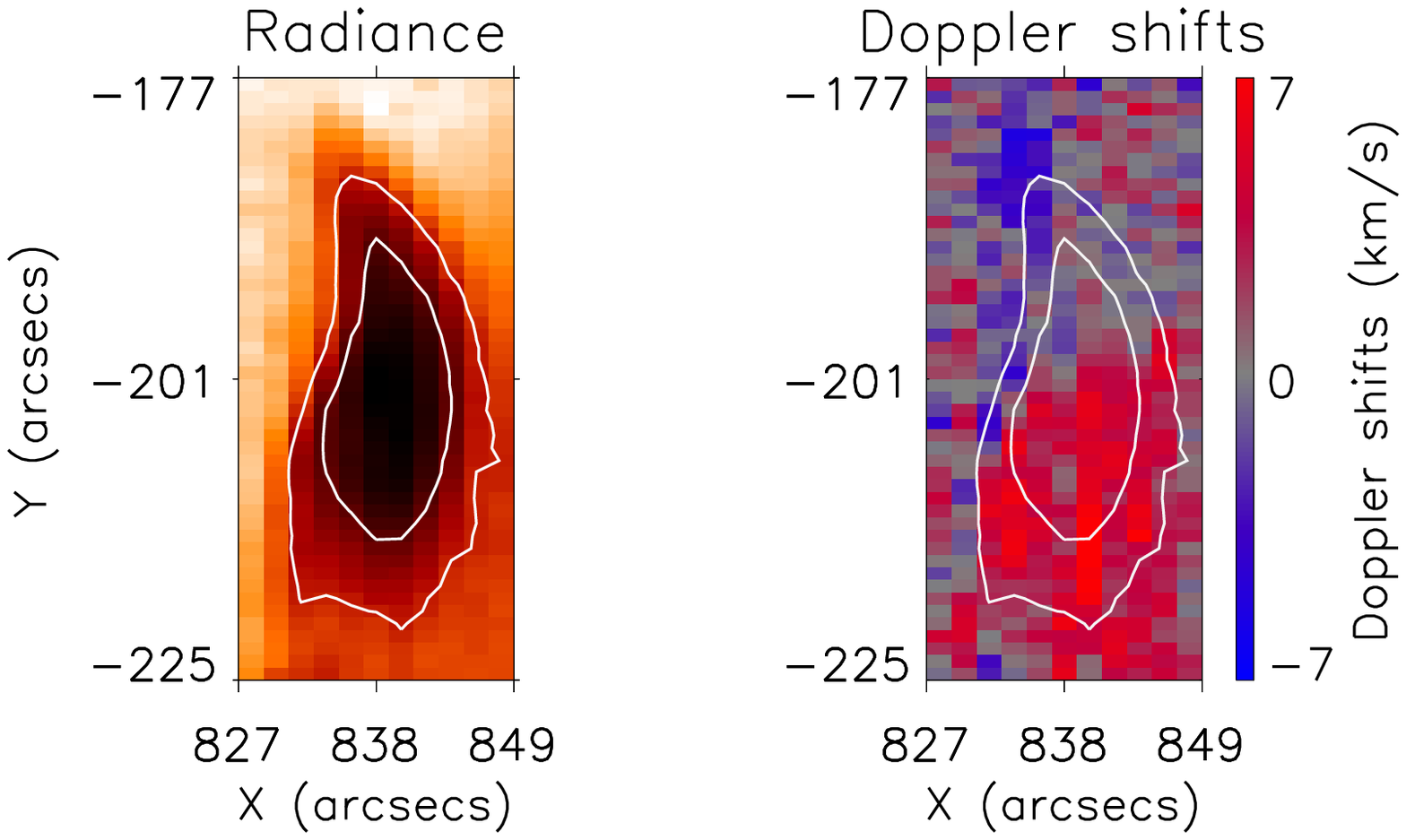}
\caption{Two examples of brightening events observed on November 16~(top: event 8 at a coronal hole boundary; bottom: event 1 inside the coronal hole). The left and right panels represent the Fe~{\sc xii}~195~\AA\ intensity  and Doppler shift images, respectively.  Event 8 was scanned from 
18:44~UT to 19:08~UT and event 1 from 18:08~UT to 18:18~UT. }
   \label{fig11}
\end{figure}

%%%%%%%%%%%%%%% Fig 12 %%%%%%%%%%%%%%%%%%%
\begin{figure}[!ht]
\centering
\includegraphics[angle=90,scale=0.45]{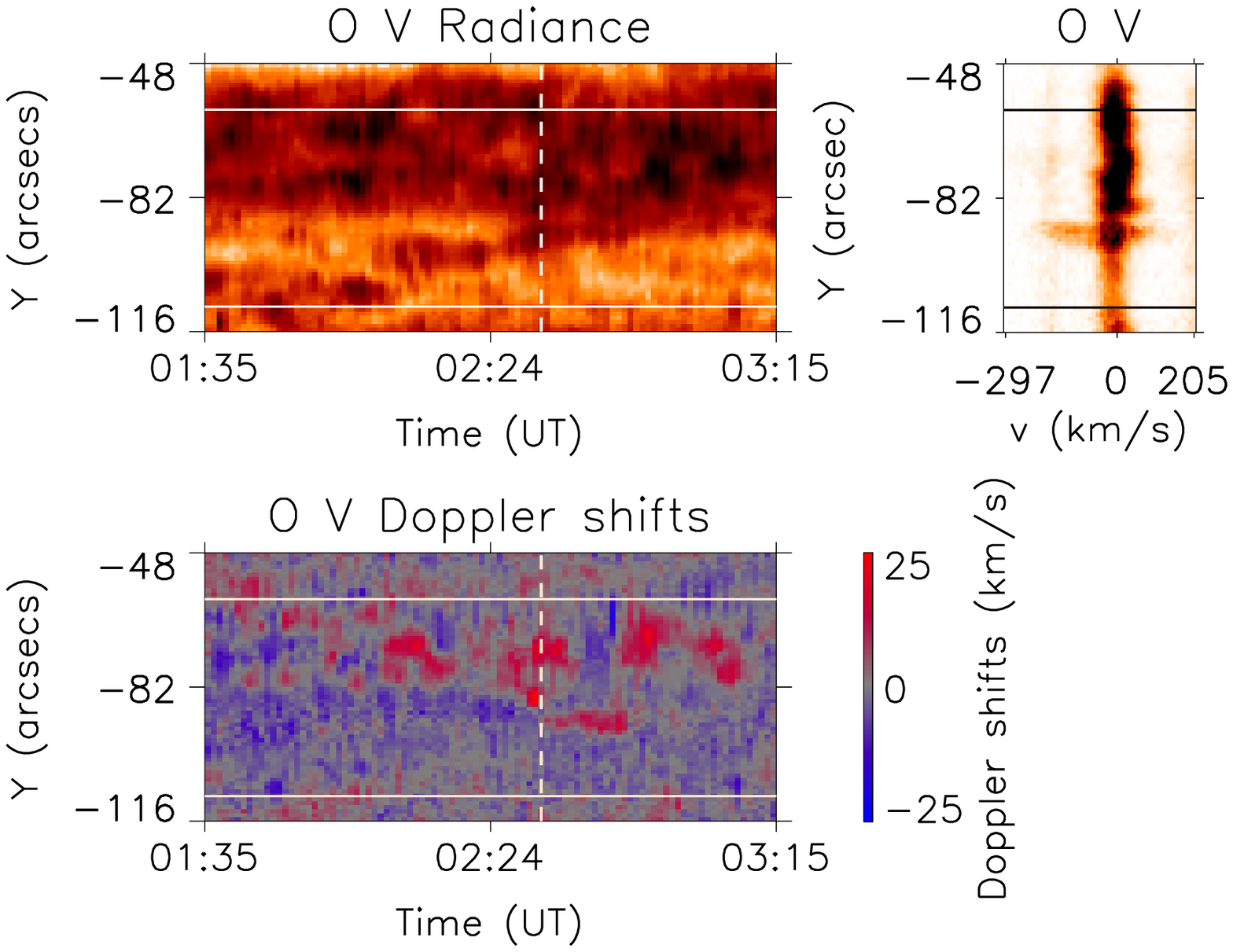}
\caption{SUMER sit-and-stare observations  of the  brightening event S2~(CHB) on  November 12.  
The radiance (top) and Doppler (bottom) maps of the event are shown obtained in the O~{\sc v} 629~\AA. The solid lines indicate the area of the event registered in X-rays. The 
last column from the top row shows the intensity along the slit at the time (02:34:35~UT) indicated with a 
dashed vertical line on the intensity image.  }
\label{fig12}
\end{figure}

In Figs.~\ref{fig8}, \ref{fig9}, \ref{fig10} and  \ref{fig11} we give examples from each day of the coronal 
hole observations. The brightening event  (No 9) shown in Fig.~\ref{fig8} (top) happens in a long-living 
large bright point at the coronal hole boundary.  It produced several outflows, two of which were significantly 
large.  The BP structure was present during the entire observing period  and it is fully in the EIS FOV only 
during the large raster. The dynamic evolution of the BP can be followed best in the SUMER sit-and-stare 
observations. The SUMER slit  lies across the multi-loop structure of the BP as shown in Fig.~\ref{fig8}.  
The evolution of the feature can be seen in the online material. The simultaneous coverage of this event 
by SUMER and EIS demonstrates once again the poor signal in the Mg~{\sc x} line while the EIS Fe~{\sc xii} 
shows a very strong increase (more on this issue see \citet{madj2010}). 
 
The brightening No. 11 (Fig.~\ref{fig3}, first panel) shown in Fig.~\ref{fig8} (bottom) represents a typical 
BP and it has the same size and shape as the QS BP shown in Fig.~\ref{fig6}. It shows a well pronounced 
half-blue and half-red Doppler shift pattern.  In the quiet phase this pattern is probably mostly due to a 
PSF effect, although, flows from one footpoint to the other are certainly present. The physical and  
instrumental effects are hard to separate. This BP shows higher density, temperature and far larger Doppler 
shifts with respect to the QS BP shown in Fig.~\ref{fig6}. The QS BP did not produce strong outflows during 
the time it was scanned by EIS. There is a small probability that  it could have had higher values for the 
plasma parameters if the measurements were taken during higher activity. However, we have to note that even 
when EIS was rastering events in the QS with strong flows, none of these events has reached the values of 
the plasma parameters of the jet producing events in CH/CHBs. 

In Fig.~\ref{fig9} we present examples of brightening caused by typical X-ray jets from coronal BPs at 
the boundary of a coronal hole (event No. 3) and inside a coronal hole (event No. 1). The event No. 3 produced 
one very faint but high velocity jet. After the jet the brightening (BP) in its footpoints disappeared 
at X-ray temperature and no more jets happened during the entire observing period. Event No. 1 started
before the beginning of our observations so we missed the time of the energy deposition but we could follow 
in full detail the X-ray jet phase. Two main flows are present, the outflow forming the jet and downflow 
in the footpoints associated with the BP. The pattern is identical to the example shown in \citet{madj2010}. 
The BP produced two more jets every time becoming smaller until it fully disappeared.

From November 14 we present two events which were in a quiet phase during the EIS scan (Fig.~\ref{fig11}). Their Doppler shifts 
are very low at coronal temperatures though the transition region O~{\sc v} line shows significant dynamics 
(see Sect.~4 for more details). The November 16 events (No. 8 and 1) are in a quiet phase during the EIS scans but 
later they produce large outflows unfortunately not detected by EIS. 

\subsubsection{Temperatures and electron densities}

Only one event from  the CH/CHB events was registered  in Fe~{\sc xxiii} (12~MK, event 4 on Nov 14) measured 
in the flaring  site (for more details see \citet{madj2010}). This event produced a large jet with a  
temperature  of maximum 2.5~MK. High temperatures of 2.5~MK (Fe~{\sc xvi}; 1 event) and 2~MK 
(Fe~{\sc xv}; 11 events) were found mostly in long-living BPs or the footpoints of large jets (e.g. event 
7 and 9 Nov 12, event 4 Nov 14). Around 62~\% (22) of the events reached only the  Fe~{\sc xiii} (1.6~MK) 
temperature. They are associated with jets, point-like brightenings, quiet phase of faint BPs (in X-rays) or 
brightenings in pre-existing loops. 
 
The average electron density of all CH/CHB events is log$_{10}{N_e}$ $\approx$9.20~cm$^{-3}$. The events which 
show electron densities above the average value  are all associated with   jet producing phenomena and their 
densities correspond to the energy deposition site, not the jets themselves. The average density of the energy 
deposition sites is log$_{10}{N_e}$ $\approx$9.51~cm$^{-3}$. Only in 5 cases the electron densities were measured 
in the jet itself (event 12 on Nov 9, events 3,6 \& 7 on Nov 12, event 3 on Nov 14) and their average electron 
density is log$_{10}{N_e}$ $\approx$8.76~cm$^{-3}$. 

%%%%%%%%%%%%%%% Fig 13 %%%%%%%%%%%%%%%%%%%
\begin{figure}[!ht]
\includegraphics[angle=90,scale=0.45]{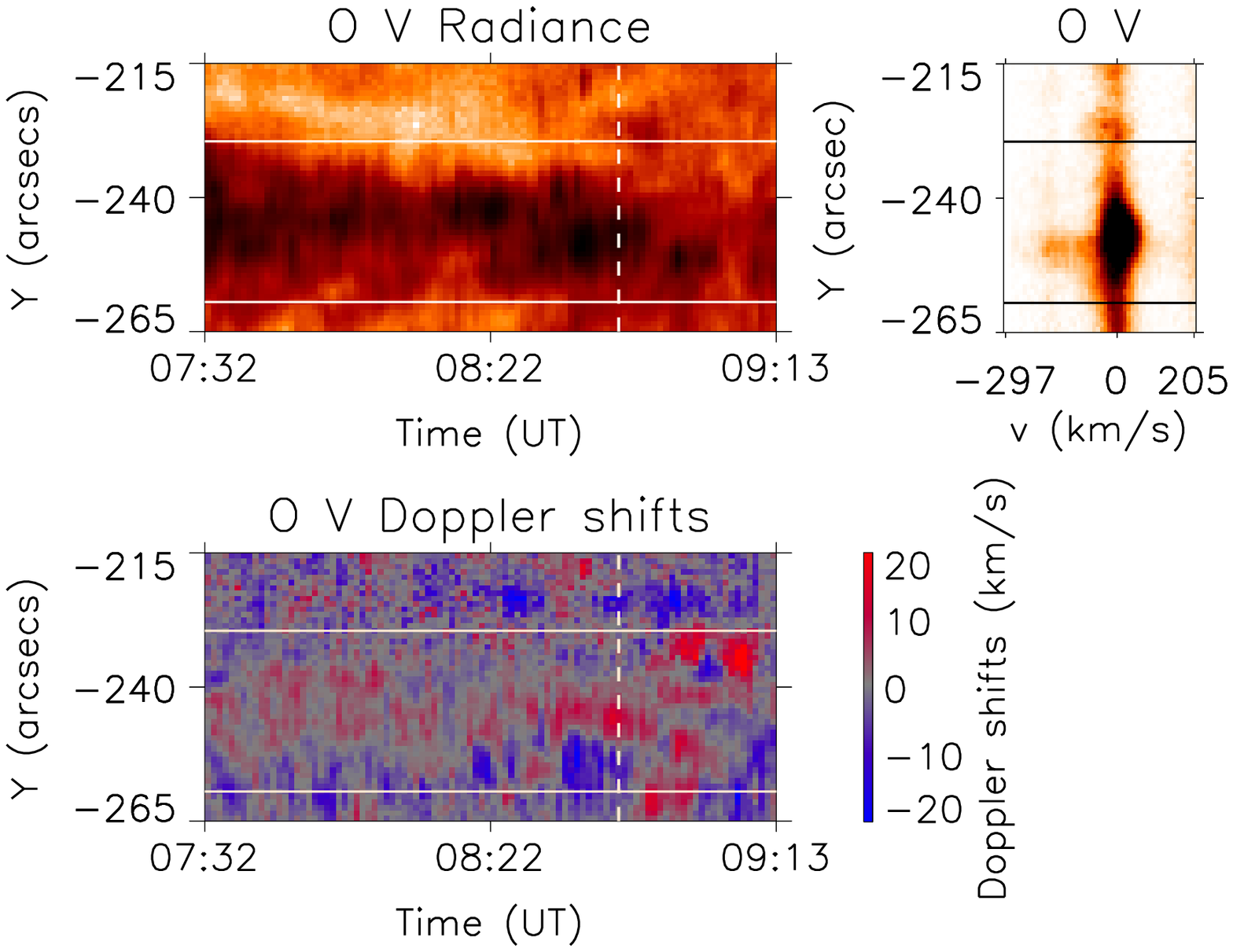}
\caption{The same as Fig.~\ref{fig12} for event S5~(CH)  on November 12. The time of the sample spectrum is 08:35:20~UT}
\label{fig13}
\end{figure}

%%%%%%%%%%%%%%% Fig 14 %%%%%%%%%%%%%%%%%%%
\begin{figure}[!ht]
\includegraphics[angle=90,scale=0.50]{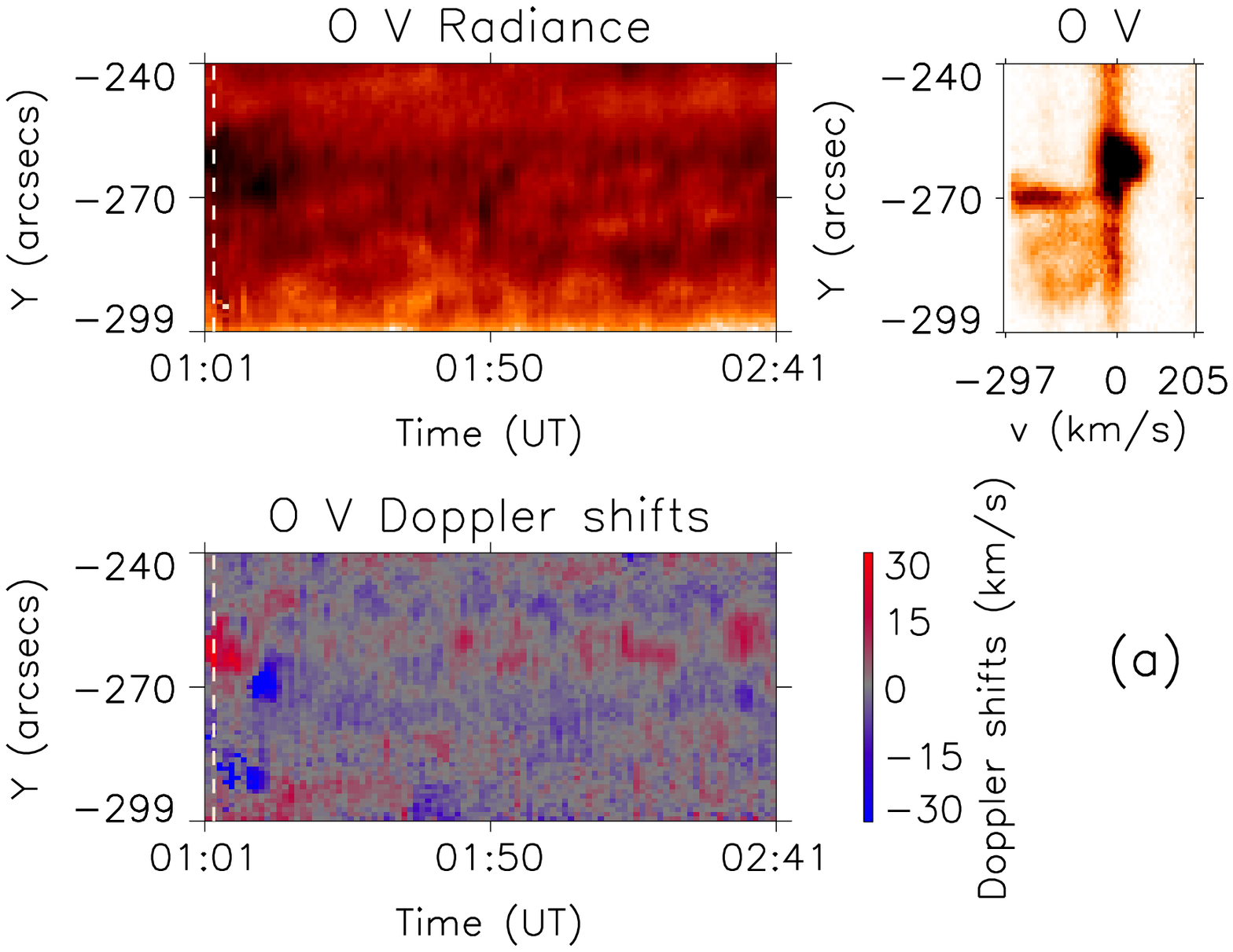}
\includegraphics[angle=90,scale=0.50]{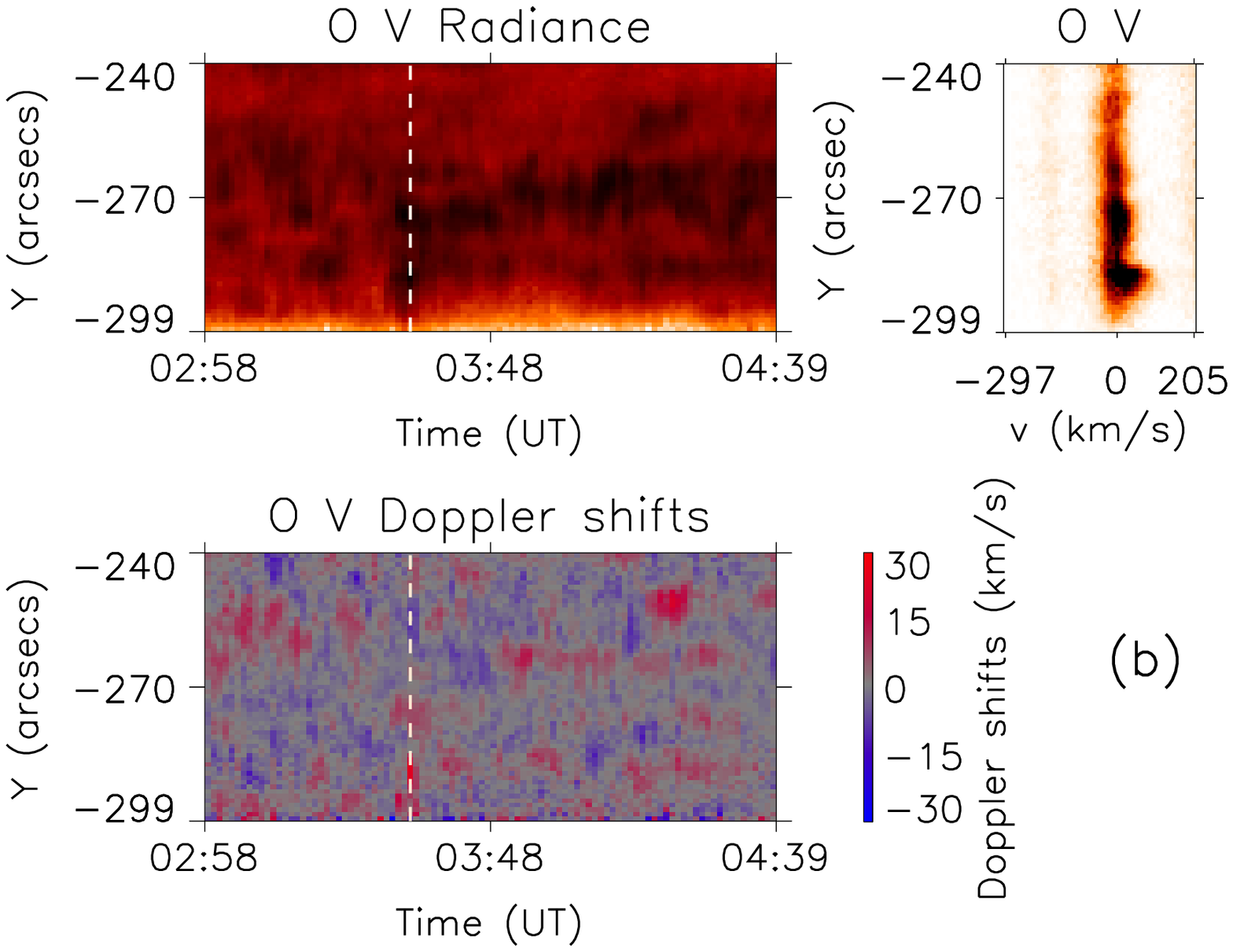}
\caption{The same as Fig.~\ref{fig12} showing event 4 (CHB) on November 14. The times of the sample spectra are 01:02:19~UT and 03:34:32~UT.}
\label{fig14}
\end{figure}

%%%%%%%%%%%%%%% Fig 15 %%%%%%%%%%%%%%%%%%%
\begin{figure}[!ht]
\includegraphics[angle=90,scale=0.50]{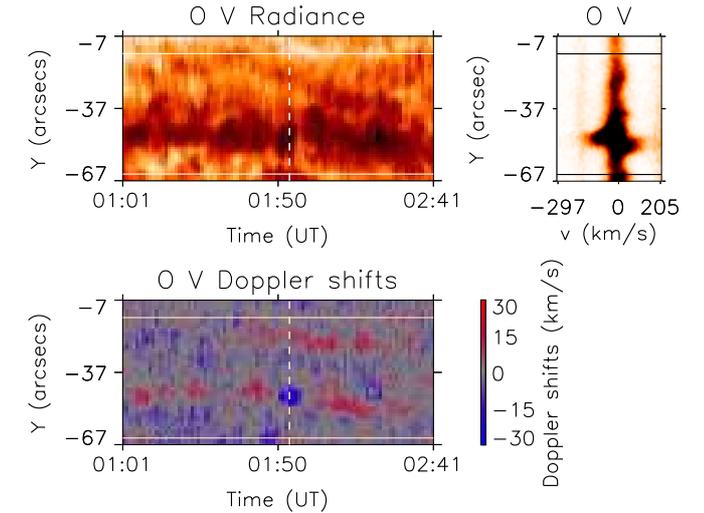}
\caption{The same as Fig.~\ref{fig12} showing event 6 (CHB) on November 14. The time of the sample spectrum is  01:55:13~UT.}
\label{fig15}
\end{figure}

%%%%%%%%%%%%%%% Fig 16 %%%%%%%%%%%%%%%%%%%
\begin{figure}[!ht]
\includegraphics[angle=90,scale=0.50]{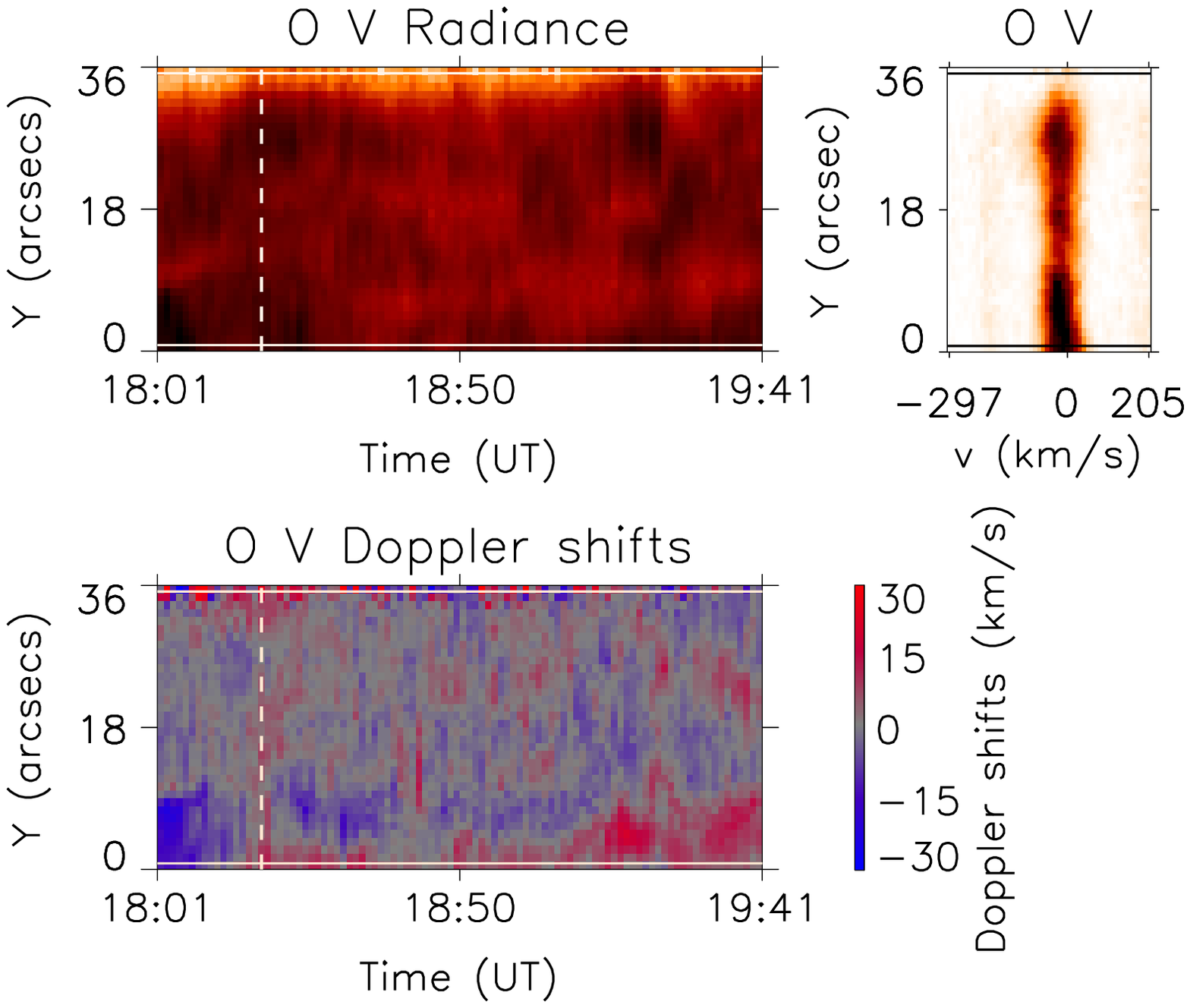}
\caption{The same as Fig.~\ref{fig12} showing event 9 (QS) on November 16. The time of the sample spectrum is  18:18:09~UT.}
\label{fig16}
\end{figure}

\section{SUMER observations of the quiet Sun and CH/CHB brightenings}

The present study provides  unprecedented analysis of  EIS and SUMER co-observations combined with 
co-temporal imaging  from XRT. Since the EIS spectrometer has very limited coverage of spectral lines with  transition 
region formation  temperatures, the simultaneous observations with SUMER contribute enormously to the understanding of the 
impact of these transient phenomena  to the dynamics and temperature gradients in the solar atmosphere.  
In Table~\ref{tb7} we present the physical parameters of the brightenings which occurred under the SUMER slit. 
The SUMER observations from November 9 suffered instrumental problems (see Sect.~2.2) and are not suitable for 
the present analysis. During part of the observations taken on November 12 the SUMER slit was pointed at a 
Solar\_x coordinate which was 96\arcsec\ East from the requested pointing. Nevertheless, we analysed these data, however, they do not have EIS coverage.  
 \begin{table*}
%\begin{center}
\centering
\caption{Physical properties of the brightening events observed by SUMER. The Mg~{\sc x} spectral window 
which also includes Si~{ii} and C~{\sc i} were transferred to the ground only when SUMER was in sit-and-stare 
(SS) mode. `R' refers to a rastering mode. The events which are not co-observed with EIS  on November 12 and 14 are 
marked with `S' followed by a digit. The co-ordinates of the events not marked in Fig.~\ref{fig1} are given in  the `Remarks' column with `(XX,YY)' in unit of arcsec.}
\begin{tabular}{c c c c c c c c c c }
\hline\hline
Date & Event No.&\multicolumn{2}{c}{Doppler shifts, O~{\sc v} (\kms)}&S~{\sc ii}/C~{\sc i}/Si~{\sc ii}&Mg~{\sc x}&Flow/Outflow\\
Identity& & Bue&Red&(response)&(response)&Remarks&\\
\hline
\hline
12/11/07&\\
CH&S1&$-$136/$-$70&16&Y&N&outflows, $>$3 repeat, SS\\
CHB&S2&$-$7&13/104&Y&N&outflows,SS; brightenings in small-scale loops\\
CHB&S3&$-$139/$-$65&28&Y&N&SS; BP\\
CH&S4&$-$85/$-$66/$-$54&23&--&--&R; small-scale brightening \\
CH&S5&$-$150/$-$80&100/24&Y&Y&outflow, $>$3 repeat, SS, (-90,-245), large active BP\\
CHB&S6&$-$50/$-$21&12&Y&N&$>$1 repeat, SS, (-100,-320), point-like brightening\\
CH&S7&$-$68/$-$34&26/17&--&--&$>$2 repeat, R, (-130,-275); short-lived small-scale BP\\
CH&1&--&70/17&--&--&R; post jet downflow\\
CHB&7&$-$150& 50 (Si~{\sc iv}) &--&--&SS\\
CH&8&$-$36& 22&--&--&R\\
\hline
14/11/07&\\
CH&3&$-$77/$-$53&--&--&--&R\\
CHB&4&$-$240& 60&Y&Y&SS\\
-- & 4a&$-$127/ -56& 7, 80\\ 
CHB&5&$-$85& 36&Y&N&SS\\
CHB&6& $-$23/$-$85& 31&Y&Y&SS\\
CHB&9&$-$64& 52&--&--&R\\
CHB&S1&$-$99& 58&--&--&R\\
\hline
16/11/07&\\
CH&5&$-$60/$-$25&--&--&--&R\\
CHB&10&$-$58&--&--&R\\
CH&11 &$-$94& 30&--&--&R\
&\\
QS&9&$-$20/$-$7&&Y&N&SS\\
AR&S1&$-$57/$-$34& 19&--&--&outflow, $>$2 repeat, R, (-705,-140)\\
\hline
\end{tabular}
%\end{center}
\label{tb7}
\end{table*}

All brightening events registered in SUMER were found to have a counterpart at  chromospheric temperatures. 
We should point out that in all cases the slit  was crossing the footpoints of the transient brightenings but 
not the jets. In a single case, event 7 on Nov 12, the slit crosses the jet  while SUMER was 
rastering in the Si~{\sc iv}~1394~\AA\ line. The chromospheric response was associated with downflows as determined from 
the transition region O~{\sc v}~629~\AA\ and N~{\sc v}~1238~\AA\ lines. 
During the  SUMER Mg~{\sc x}~625~\AA\  and EIS Fe~{\sc xii}~195~\AA\  co-observations, only a third of the events had a response in the Mg~{\sc x}~625~\AA\ line while all events had a signature in the Fe~{\sc xii}~195~\AA\ line.
Discussion on the Mg~{\sc x} response can be found in \citet{madj2010}. A 
significant part of the transition-region dynamics seen as Doppler shifts in the O~{\sc v} line has no counterpart 
at X-ray temperatures. We believe that spicules may be mostly responsible for this dynamics 
\citep{1983ApJ...267L..65D,2003A&A...403..731M}. We intentionally did not include the analysis of the observations 
taken in the Si~{\sc iv} line. This line is formed at temperatures of 80\,000~K and  would detect  the
upper-chromosphere/low-transition-region spicule activity. Therefore, it is 
difficult and often impossible to separate a line profile produced by events originating in the chromosphere from events which originate in the corona and leave their imprint in the transition region. We, however, included the 
Si~{\sc iv} line in the animation sequence for the purpose of continuity of the presentation. From the comparison 
of the Doppler shift measurements shown in Table~\ref{tb7} with the measurements in the Fe~{\sc xii} line given in 
Table~\ref{tb6}, we can clearly see that transient brightenings triggered by X-ray jets or just flows in loops 
contribute significantly to the dynamics of the transition region. Flows are seen at all temperatures from 
the chromosphere to the corona which demonstrates the impact on the mass flow in the solar atmosphere of events 
associated with transient brightenings. We give a few typical examples of events seen in the SUMER O~{\sc v} line and discuss their 
identification with well know phenomena from past spectrometer observations (HRTS \& SUMER).

Typical examples of the SUMER O~{\sc v} line profiles produced by the transient brightenings associated with jets or 
flows in loops can be seen in Figs.~12 to 16. The event S2 which is only seen in SUMER and XRT, evolved as a point-like 
brightening which produced  faint X-ray jet(s) (the jet appearance is so faint in X-rays that it is difficult to judge 
how many jets have been produced). In response to this phenomenon, the O~{\sc v} line showed strong blue- and 
red-shifted emission  typical for explosive events (Fig.~\ref{fig12}). The Doppler shift appeared in a burst of a 
little more than 20~min often associated with EEs. 

After the slit moved to a position which was 96\arcsec\ from the originally planned, it stared at the edge of a 
large dynamically evolving BP. In Fig.~\ref{fig13} we show the intensity and Doppler shift  images of this phenomenon. 
Again, dynamically evolving emission in the wings of the O~{\sc v} line appear in all shapes associated with jets, 
flows or explosive events (see the online material  Fig.~A\ref{fig18}). 

The next example given in Fig.~\ref{fig14} is the event studied in detail by \citet{madj2010}. We show here the 
reoccurrence of the jet in the same BP (mentioned but not discussed in \citet{madj2010}). During the first jet the 
slit was positioned on the propagation path of the jet, while during the second jet, the slit crossed the opposite 
edge of the BP. The first jet (Fig.~\ref{fig14}a) is seen as a very strong blue-shifted emission and the downflow as 
a red-shifted emission  in the O~{\sc v} line. At the time of  the second jet  only the downflow can be seen 
(Fig.~\ref{fig14}b)  as the SUMER slit does not stare at the jet  (see the online material for more details).  

The example given in Fig.~\ref{fig15} occurred at the coronal hole boundary. The event is hardly seen evolving 
during the EIS  big image scan. In the SUMER sit-and-stare observations, however,  the dynamics is clearly visible as 
varying blue- and red-shifted emission (see the November 14 animation for more details). 

The last example given in Fig.~\ref{fig16} is the only quiet Sun event recorded in the SUMER data. Brightenings 
along the SUMER slit with blue and red-shifted emission which do not exceed 50~\kms\ are seen during the event 
(see the November 16 animation for more details).

\section{Discussion and Conclusions}

Knowledge on the plasma properties of small-scale transient phenomena is crucial for their modelling as well as the understanding of their 
role in coronal heating and solar wind generation. The present paper delivers a follow-up spectroscopic study on a large 
number of transient small-scale events in coronal holes and their boundaries.  The counterparts of these phenomena in the quiet Sun were also analysed.

During the past decade  several 
groups \citep{2000ApJ...542.1100S, 2007PASJ...59S.751C, 2007PASJ...59S.771S,2010ApJ...710.1806D}. \citet{2000ApJ...542.1100S} investigated 
the physical parameters of coronal jets mostly known as X-ray jets following their discovery in images from SXT/Yohkoh. 
\citet{2010ApJ...710.1806D} obtained electron  density of log$_{10}{N_e}$ 
$\approx$8.85~cm$^{-3}$ at the base of a jet using the same line ratio as in the present study. The electron density of 
jets log$_{10}{N_e}$ $\approx$8.76~cm$^{-3}$ measured here is very close to their value. Furthermore, the present study provides 
measurements of the electron density in the flaring site which is on average  log$_{10}{N_e}$ $\approx$9.51~cm$^{-3}$. We derived 
similar differences in the electron density measured in the microflaring and flow/outflow sites in quiet Sun transients which strongly 
suggests that the same physical mechanism may trigger small-scale transients anywhere in the solar atmosphere.  \citet{2000ApJ...542.1100S} 
obtained the electron densities 
of 16 jets using the soft X-ray imager onboard Yohkoh. They measured electron densities of  
log$_{10}{N_e}$ $\approx$9.73~cm$^{-3}$ in the microflaring sites and log$_{10}{N_e}$ $\approx$9.23~cm$^{-3}$  for the jets.
These values are 60\% higher for the energy deposition site and almost two times higher for the  jets in comparison to the values obtained here. The differences probably come from the filling factor assumptions with the imager data. We should  also bear in mind that spectroscopic 
measurements of electron densities using spectral  lines from the same element and ionization stage is the most 
reliable method for density diagnostics.

The present study uses the widest temperature range coverage of small-scale transients thanks to the combination of 
SUMER and EIS observations, e.g. from 10\,000~K to 12~MK. \citet{2007PASJ...59S.751C} found that the temperatures of BPs  and jets are not higher than 2.5~MK (EIS slot observations) while \citet{2010ApJ...710.1806D} (EIS slit observations) obtained temperatures from 0.3 
to 2.2~MK. \citet{2000ApJ...542.1100S} found similar temperatures of 3--6~MK for both the jets and flaring sites.
Here, the jet temperatures reached a maximum of 2.5~MK but in the majority of the cases the 
temperatures did not exceed 1.6~MK (Fe~{\sc xiii}). The footpoints of jets which present large 
and long-lived coronal bright points seen in X-ray images had maximum temperatures of 2.5~MK.  The temperature  of the flaring site of one event reached 
12~MK.   Only during this jet did  the slit cross the flaring site a minute after the 
energy deposition took place. We believe that the higher temperature in the footpoints of jets must be a common property, but 
to confirm this additional observational evidence is needed. No obvious difference was found between the temperatures of the quiet 
Sun and coronal hole transient events, pointing again towards the same physical
mechanism for the origin.

The spectroscopic analysis of small-scale transient brightenings indicates that the majority of the phenomena reaching 
high coronal  temperatures were triggered by multiple microflarings (during a single event) which take place in
either  the transition region or the corona, but no indication was found for energy deposition in the chromosphere. Chromospheric emission was mostly related to downflows in the footpoints of jets. 

 The similar temperatures and densities of the quiet Sun and coronal hole events, as well as the same observational signature of the  
 onset of these events, i.e. microflaring, agree with the suggestion by  \citet{1995Natur.375...42Y} that the 
physical origin of both jets (CH events) and loop brightenings (typical for the QS) are possibly the same, i.e. 
magnetic reconnection.  The only apparent difference is the higher dynamics of the CH/CHBs events in terms of high velocity plasma flows. 
This can be explained via the different magnetic field configuration involved. A reconnection of open and closed magnetic field lines 
in the CHs and at CHBs will result in the  acceleration of the ejected plasma to higher speeds due to the lower density along the open field lines. 

We can conclude that the dynamic day-by-day and even hour-by-hour small-scale evolution of coronal hole boundaries reported in paper~I 
is indeed related to coronal bright points. The XRT observations reported in paper~II revealed that these changes result from  
the dynamic evolution of coronal bright points producing multiple jets during their lifetime until their full disappearance.  
\citet{2004ApJ...603L..57M}  first detected the spectroscopic signature  of these dynamic changes. This signature, in the form 
of non-Gaussian profiles of spectral lines at transition region temperatures, was believed to be produced by the so-called `explosive 
events' (see Sect.~1 for details) with the strong blue- and red-shifted emission resulting from bi-directional jets expelled from a 
current sheet due to magnetic reconnection. The present study  added the last pieces  to the puzzle by revealing what 
phenomena do produce these bi-directional jets, their origin, evolution, temperatures etc. We found that numerous jets from coronal 
bright points in coronal holes and at their boundaries  do carry high density plasma at very high speeds  throughout  the solar atmosphere.  
It is now known that large numbers of these jets reach the interplanetary space \citep{2010SoPh..264..365P}.
More recently, Jackson et al. (during the SDO-4/IRIS/Hinode workshop 2012) reported their finding of  ``a positive correlation between the brightest of the polar 
jets and a high-speed response traced into the interplanetary medium''.

The concentration of jet-like events in  coronal holes and at their boundaries as shown in paper~II  and their plasma characteristics obtained  here 
make them an important candidate for one of the sources of the slow solar wind. Most recently, though, \citet{2012ApJ...750...50N} 
`proposed that the interplanetary manifestations of X-ray jets observed in solar polar coronal holes during
periods of low solar activity are the peaks of the so-called microstreams observed in the fast polar solar wind'. 
We already know that there is no physical difference between equatorial and polar jets, therefore, the possibility that 
X-ray jet propagate into the fast solar wind as microstreams should not be excluded. 

The present study demonstrates the association of X-ray jets and flows in loops as one of the main phenomena 
which produces 'explosive events'. \citet{2002A&A...392..309T} suggested that explosive events or rather events with strong non-Gaussian profiles 
 do not reach coronal temperatures based on their response in the SUMER Mg~{\sc x} line. Note that the study was based on a few events only.
 The present study shows clearly that this line is not reliable for such diagnostics 
(see detail discussion on this in \citet{madj2010}). Therefore, the possibility 
that the events discussed here may have a significant contribution to the temperature gradient in the solar 
atmosphere should be reconsidered. This subject needs to be revisited in the light of the results presented here, and new statistical 
studies based on imaging data (e.g. AIA on the Solar Dynamic Observatory) supported by spectroscopic 
information should re-evaluate the power distribution of these events. 
Recently, a multi-instrument study by \citet{2012A&A...538A..50S} of another phenomena know as blinkers or transition 
region brightenings,  which were known only by their appearance in spectroscopic data (mainly from the Coronal 
Diagnostics Spectrometer onboard SoHO) showed that they cover a wide range of phenomena. Blinkers were found to be the EUV response of 
various transient
events originating at coronal, transition region and chromospheric heights as are the 
explosive events. Hence, both blinkers and explosive events will contribute to the
formation and maintenance 
of the temperature gradient in the transition region and the corona. 

This  study demonstrates the immense 
capabilities of the present imaging and spectroscopic instrumentation. It also shows how in the future the 
forthcoming IRIS (Interface Region Imaging Spectrograph)  instrument can be successfully combined with the already existing 
imaging and spectroscopy instruments like EIS/SOT/XRT/Hinode and AIA/HMI/SDO. The next part of this extensive 
study will present  the magnetic field evolution  (SOT/Hinode) associated with the  small-scale transients discussed here. 

\begin{acknowledgements}  
We thank very much the anonymous referee for the important correction and sugegtsions. 
Research at Armagh Observatory is grant-aided by the N.~Ireland Department of Culture, Arts and 
Leisure. We also thank STFC for support via grant ST/J001082/1. Hinode is a 
Japanese mission developed and launched by ISAS/JAXA, with NAOJ as domestic partner and NASA and 
STFC (UK) as international partners. It is operated by these agencies in co-operation with ESA 
and NSC (Norway). The research leading to these results has received funding from the European 
Commission's Seventh Framework Programme (FP7/2007-2013) under the grant agreement eHeroes 
(project n¡ 284461, www.eheroes.eu).The authors MM and JGD thank ISSI, Bern for the support of the team 
``Small-scale transient phenomena and their contribution to coronal heating''.

\end{acknowledgements}

\bibliographystyle{aa}

\Online
\begin{appendix}
\begin{figure*}
\centering
\includegraphics[width=15cm]{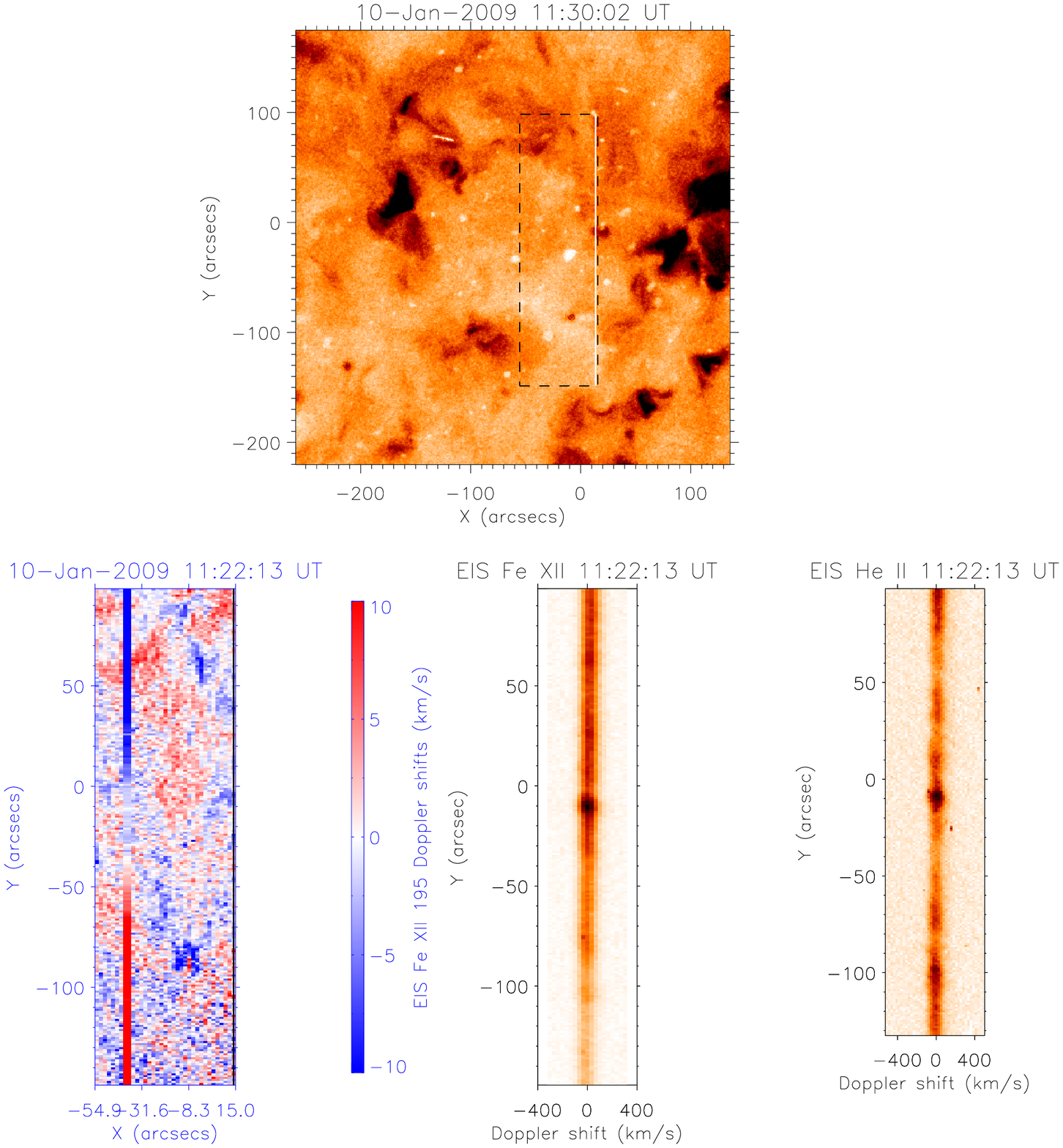}
\vspace{-1cm}
\caption{Animation of co-temporal images of the quiet Sun dataset on January 10 including XRT images on top with overplotted the EIS raster FOV with dashed lines and solid white line the EIS slit at the given time. Bottom from left to right: EIS Doppler maps, Fe~{\sc xii} 195.12~\AA\  and He~{\sc II}~256.32~\AA\ intensity along the slit.  }
\label{fig17}
\end{figure*}

\begin{figure*}
\centering
\includegraphics[width=15cm]{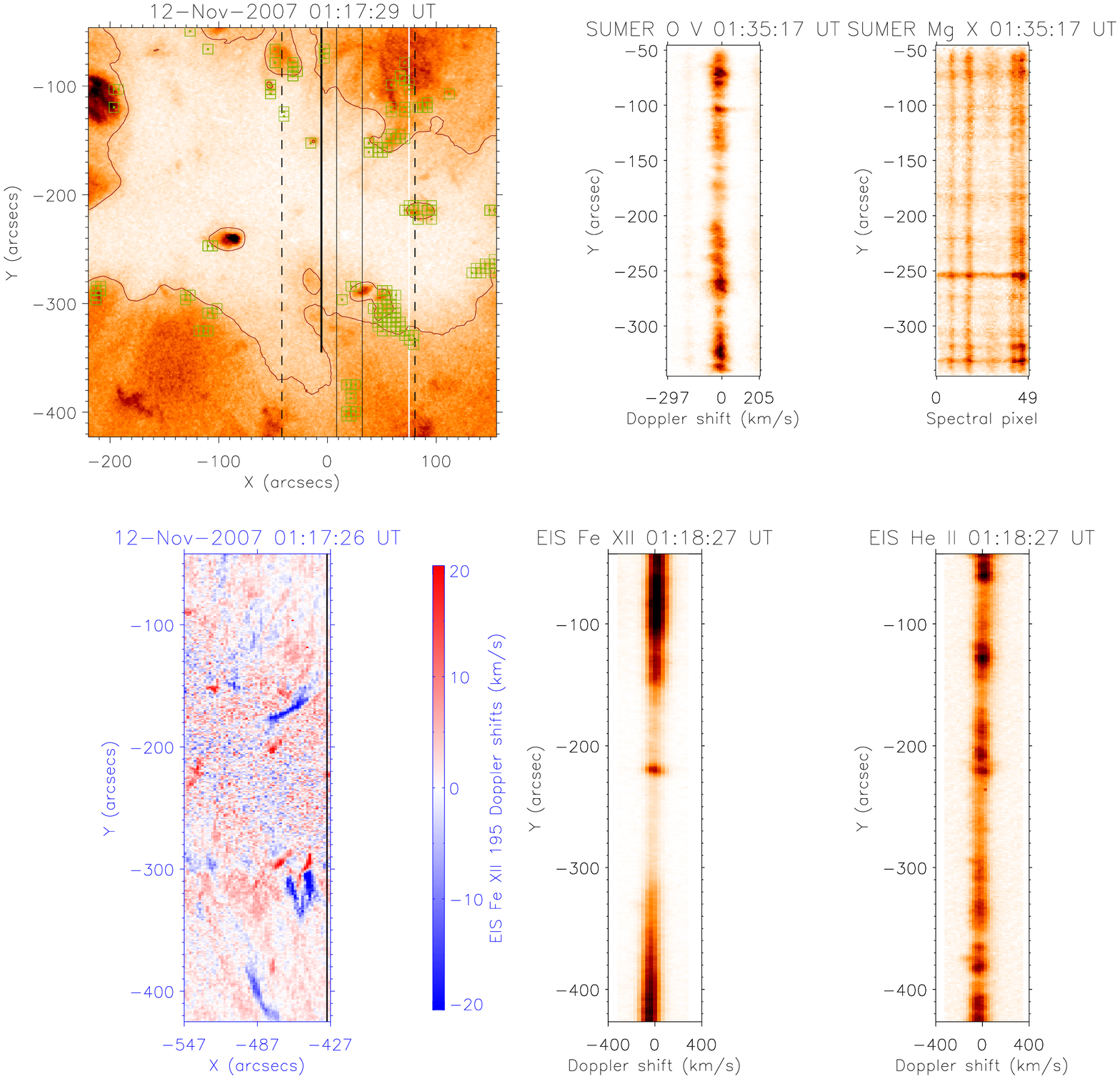}
\vspace{-1cm}
\caption{Animation of co-temporal images of the quiet Sun dataset on November 12 including on top XRT images with overplotted the EIS big raster FOV with dashed lines, solid long lines the EIS small raster and solid short line the SUMER slit position at the given time, and intensity of SUMER O~{\sc v}~1259.54~\AA\ and SUMER Mg~{\sc x}~1249.90~\AA\ windows. The overplotted  white line represents the EIS slit  position at the given time.   Bottom from left to right: EIS Doppler maps, Fe~{\sc xii} 195.12~\AA\  and He~{\sc II}~256.32~\AA\ intensity along the slit at the given time. }
\label{fig18}
\end{figure*}

\end{appendix}

\end{document}